

\def\cl#1{\centerline{#1}}
\def\sv#1{\vskip#1ex}
\def\VE{\vfill\eject}
\def\ie{\sl i.e., \rm}
\def\eg{\sl e.g., \rm}
\def\st{\hbox{\ \rm such that \ }}
\def\qed{\vrule height6pt width3pt depth 0pt} 
\def\pf{\sl Proof: \rm} 
\def\ZZ{{\bf Z}}
\def\JMP{\sl J. Math. Phys. \rm}

\def\skp{\vskip 1ex}
\def\n{\|\kern-.03cm |}

\def\cc{{\bf C}}

\def\cn{{\cc^n}}

\def\rr{{\bf R}}
\def\rh{{\widehat\rr}}
\def\rs{{\widehat\cc}}
\def\rn{{\rr^n}}
\def\rd{{\rr^d}}

\def\x{{\bf x}}
\def\y{{\bf y}}
\def\v{{\bf v}}
\def\p{{\bf p}}
\def\q{{\bf q}}
\def\k{{\bf k}}
\def\e{{\bf e}}
\def\t{{\bf t}}
\def\o{{\bf 0}}
\def\b{{\bf b}}
\def\z{{\bf z}}
\def\h{{1\over 2}}
\def\l{\langle\, }
\def\r{\,\rangle}
\def\L{\langle\kern-.06cm\langle\, }
\def\R{\rangle\kern-.06cm\rangle}
\def\r{\,\rangle}
\def\inr{\int_{-\infty}^\infty}
\def\hh{\hat h}
\def\fh{\hat f}
\def\fpp{f_+^+}
\def\fpm{f_+^-}
\def\fmp{f_-^+}
\def\fmm{f_-^-}
\def\epu{e^{2\pi ipu \,}}
\def\epv{e^{2\pi ipv \,}}
\def\ez{{e_z}}
\def\he{\hat e_z{}}
\def\ph{\hat\phi  }

\def\ftil{\tilde f}
\def\gtil{\tilde g}
\def\ca{{\cal A}}
\def\ch{{\cal H}}
\def\hs{{\ch_s}}
\def\cd{{\cal D}}
\def\ds{{\cd_s}}
\def\cp{{\cal P}}
\def\inds{{ 1\over 2\pi i}\int_{-\infty}^\infty\,{ d\tau \over\tau -i}\,} 
\def\inp{\int_{\rn} d^n\p\,\,}
\def\eyp{\theta (\p\cdot \y)\,e^{2\pi i\p\cdot \z}}
\def\del{{\sqcup\hskip-1.6ex\sqcap\ \!\!}}
\def\dir{\partial\kern -1.2ex /\,}
\def\tb{{\cal T}}

\def\om{{\Omega _m}}
\def\omp{{\Omega _m^+}}
\def\omm{{\Omega _m^-}}

\def\lt{L^2(\rr)}
\def\slt{SL(2, \rr)}


\magnification 1200
\parindent=0pt

\cl{\bf Windowed Radon Transforms, Analytic Signals and the Wave Equation} 

\sv2

\cl{Appeared in \sl Wavelets: A Tutorial in Theory and Applications}
\cl{C K Chui, Editor, Academic Press, 1992}
\sv2

\centerline{\bf G.~Kaiser} 
\centerline{Department of Mathematics}
\centerline{Univ. of Massachusetts, Lowell, MA 01854, USA} 
\bigskip
\centerline{\bf R.~F.~Streater} 
\centerline{Department of Mathematics} 
\centerline{King's College, London WC2R 2LS, England} 
\sv1

\centerline{October 1991}

\sv5

\cl{\bf Abstract} 
\sv1

The act of measuring a physical signal or field suggests  a 
generalization of the  wavelet transform that  turns out to be a windowed
version of the Radon transform.  A reconstruction formula is derived which
inverts this transform. A special choice of window yields  the
``Analytic--Signal transform'' (AST), which gives a partially analytic
extension of functions from $\rn$ to $\cn$.  For n =1, this reduces to Gabor's
classical definition of ``analytic signals.''   The AST is applied to  the
wave equation, giving an expansion of solutions in terms of wavelets
specifically adapted to that equation and parametrized by real space and
imaginary time coordinates (the  {\rm Euclidean region)}.

\sv4

\cl{\bf 1. Introduction}
\sv1

\noindent The ideas presented here originated in relativistic quantum 
theory [13, 14, 15], where a method was developed for extending arbitrary
functions from $\rn$ to $\cn$ in a semi--analytic way.  This gave rise to the
``Analytic--Signal transform'' (AST) [16].  Later it was realized that the AST
has a natural generalization to what we have called a Windowed X--Ray
transform [17], and the latter is a special case of a Windowed  Radon
transform, to be introduced below.  For $n=1$, these transforms reduce to the
(continuous) Wavelet transform.  In the general case, they retain many of
the properties of the Wavelet transform.

 \skp

In Section 2 we motivate and define the d--dimensional Windowed Radon
transform in $\rn$ for $1\le d\le n$ and derive  reconstruction formulas
which can be used to invert it.  In Section 3 we define the AST in $\rn$
and give some of its applications.  In Section 4 we develop a new
application of the AST by generalizing a construction in [16] to the wave
equation in $\rr^2$. This results in a representation  of solutions of the wave
equation  as combinations of  ``dedicated'' wavelets that
 are especially customized to that equation.  In
particular, these wavelets are themselves solutions  and
represent {\sl coherent wave packets,\/}  being well-localized in both space
(at any particular time) and frequency, within the limitations of the
uncertainty principle.  The parameters labeling these wavelets (\ie the
variables on which the AST depends) have a direct geometrical significance: 
They give the initial position, direction of motion and average frequency or 
{\sl color\/}  of the wavelets.   The representation  of  a solution in terms of
these wavelets therefore gives a {\sl geometrical--optics\/}  (ray)  picture of
the solution.   It is suggested that this could be of considerable practical
value in signal analysis, since many  naturally occurring  signals (\eg sound
waves, electromagnetic waves) satisfy the wave equation away from sources
and the geometrical--optics picture gives a readily accessible display of their
informational contents.

\sv4

\cl{\bf 2. Windowed Radon Transforms}
\sv1

\bf 2.1. The  Windowed X--Ray Transform \rm
\sv1

\noindent Suppose we wish to measure a physical field distributed in $\rn$.  
This field could be a ``signal,'' such as an electromagnetic field 
or the pressure distribution due to a sound wave.  For simplicity, we assume
to begin with that it is real--valued, such as pressure.  (Our considerations
easily extend to  complex--valued, vector--valued or tensor--valued signals,
such as electromagnetic fields; we shall indicate later how this is done.)  The
given field is therefore a function $f:\/ \rn\to\rr$.  We may think of $\rn$ as
physical space (so that $n=3$), in which case the field is time--independent,
or as space--time (so that $n=4$), in which case the field may be
time--dependent. In the former case, $\rn$ is endowed with a Euclidean
metric, while in the latter case the appropriate metric is Lorentzian, as
mandated by Relativity theory. 
\skp

Actual measurements are never instantaneous, nor do they take place at
a single point in space.  A measurement is performed by reading an
instrument, and the instrument necessarily occupies some region in space
and  must interact with the field for some time--interval before giving a
meaningful reading.  Let us assume, to begin with, that the spatial extension
of the instrument is negligible, so that it can be regarded as being
concentrated at a single point at any time.  We allow our instrument to be in
an arbitrary state of uniform motion, so that its position is given by
$\x(t)=\x+\v t$, where $t\in\rr$ is a ``time'' parameter and $\x,
\v\in\rn$.  Note that $t$ need not be the physical time.  For example, if
$\rn$ is space--time, then each ``point'' $\x$ represents an {\sl event,\/}
i.e. a particular location in space at a particular time.  In that case,  the line
$\x(t)$ is called a {\sl world--line\/}  and represents the entire history of
the point--instrument.  The ``velocity'' vector $\v$ then has one too many
components and may be regarded as a set of {\sl homogeneous
coordinates\/}  for the physical velocity.   Note that in this case  $\v$
cannot vanish, since this would correspond to an instrument not subject to
the flow of time.  Even if $\rn$ is space, the case $\v=\o$ is not interesting
since then the instrument can only measure the field at  a single point.  We
therefore assume that $\v\ne\o$, hence  $\v\in\rr^n_*\equiv
\rn\backslash\{\o\}$.

\skp

Let us assume that the reading registered by the  instrument  at time $s$
gives a weight $h(t-s)$ to the value of the field passed by the instrument  at
time $t$.  (For motivational purposes we note that causality would demand
that $h(t-s)=0$ for $t>s$;  moreover, $h(t-s)$ should be concentrated in some
interval $s-\tau \le t\le s$, where $\tau $ is a ``response time'' or memory
characteristic of the instrument.  However, the results below do not depend
on these assumptions.)  Our model for the observed value of the field at the
``point'' $\x$, as measured by the instrument traveling with uniform velocity
$\v$, is then

$$
f_h(\x, \v)\equiv \inr dt\,h(t) f(\x+\v t). 
\eqno(1)
 $$

To accomodate complex--valued signals, we allow the weight
function $h$ to be complex--valued.   $h(t)^*$ will denote the complex
conjugate of $h(t)$.  In order to minimize analytical subtleties, we assume
that $h$ is smooth and bounded, and that $f$ is smooth with rapid decay
(say, a Schwartz test function).

\proclaim Definition 1.  The Windowed X--Ray Transform  of $f:\/
\rn\to\cc$ is the function $f:\/ \rn\times\rr^n_*\to\cc$ given by
$$
 f_h(\x, \v)=\inr dt\,h(t)^*f(\x+\v t).
\eqno(2)
 $$

\skp

\noindent {\sl Remarks.\/}  

\item{1. } In the special case $h(t)\equiv 1$ and $|\v|=1$, $f_h$ is known
as the (ordinary) {\sl X--Ray transform\/} of $f$ (Helgason [11]), due to its
applications in tomography.  We may then regard $f_h$  as being defined on
the set of all lines in $\rn$,  independent of their parametrization.  In the
general case, we think of the function $h(t)$ as a {\sl window,\/} which
explains our terminology.

\smallskip

\item{2.} Some  work along related lines was recently done by
Holschneider [12].   He considers a two--dimensional wavelet transform
which is covariant under translations, rotations and and dilations of $\rr^2$.
When the window function  is supported on a line, say $h(t_1, t_2)=\delta
(t_2)$, this becomes an X--Ray transform in $\rr^2$.  His inversion
method is less direct than ours in that it involves a limiting process.

\smallskip

\item{3.}  Note  that $f_h$ has  the following {\sl dilation property\/}
for $a\ne 0$:
$$
 f_h(\x, a\v )=\inr dt\,|a |^{-1} \, h(t/a)^*\,f(\x + t\v ) = 
f_{h_a}(\x, \v ) 
\eqno(3)
 $$
where $h_a (t)\equiv |a|^{-1} \,h(t/a)$.  This may be used to study the
behavior of $f_h$ as $\v\to\o$.   For the ``forbidden'' value $\v=\o$, 
the transform becomes  $f_h(\x, \o)=\hh(0)^*f(\x)$, where $\hh$ is the
Fourier transform of $h$.  (We shall see that $\hh(0)=0$ for ``admissible''
$h$.) 

\smallskip

\item{4.} For $n=1$ and $v  \ne 0$, a change of variables
gives 
$$\eqalign{
f_h(x,v) & =|v|^{-1} \inr dt'\, h\left({ t'-x\over v}  \right)^*\,f(t')\cr
 & =|v|^{-1/2}\,Wf(x,v),
 \cr}
 \eqno(4)
 $$
where $Wf$ is the usual wavelet transform of $f$ [5, 7, 23], with $v$ playing
the role of a {\sl dilation factor\/}    and the window function $h(t)$  playing
the role of a {\sl basic wavelet.\/} 

\smallskip

\item{5.}  All our considerations extend to vector--valued signals.  The
cleanest approach is to let the window function $h^*$ assume values in the
{\sl dual\/}  vector space, so that $h(t)^*f(\x+\v t)$ and $f_h(\x, \v)$ are
scalars.  More than one window needs to be used (or rotated versions of a
single window), in order to `probe' the different components of $f$.  The
same applies to tensor--valued signals such as electromagnetic fields, since
they may be regarded as being valued in a higher--dimensional vector
space.   However, a more correct way to measure a vector-- or
tensor--valued field is to use an instrument which is not rotationally
invariant, and that implies that the instrument has some spatial extension. 
This is done in Section 2.3.

\skp

It will be useful to write $f_h$ in another form 
by substituting the Fourier representation  of $f$
into $f_h$.  Formally, this gives

$$\eqalign{
f_h(\x,\v) & =\inr dt\,\int_\rn d\p\,e^{2\pi i\p\cdot(\x+t\v)}\,
h(t)^*\,\fh(\p)\cr
 & =\int_\rn d\p\,e^{2\pi i\p\cdot\x}\,\hh(\p\cdot\v)^*\,\fh(\p)\cr  
 & \equiv \langle\, \hh_{\x,\v}\,  , \fh\,\rangle_{L^2(d\p)}=
\langle\,  h_{\x,\v}\,  , f\,\rangle_{L^2(d\x)},
 \cr}
 \eqno(5)
 $$ 
where $\hh_{\x,\v}$ is defined by

$$
 \hh_{\x,\v}(\p)=e^{-2\pi i \p\cdot\x}\,\hh(\p\cdot\v), 
\eqno(6)
 $$
so that 

$$
 h_{\x,\v}(\x')=\int_\rn d\p\,e^{2\pi i\p\cdot(\x'-\x)}\,\hh(\p\cdot\v).
\eqno(7)
 $$
(We have adopted the convention used in the physics literature, where
complex inner products are linear in the {\sl second\/}  factor and
antilinear in the first factor.)  

The functions $ h_{\x,\v}$ are 
$n$--dimensional ``wavelets'' and will be used in the next subsection to
reconstruct the signal $f$.   Note that $\hh_{\x,\v}$ (hence also
$h_{\x,\v}$) is not square--integrable for $n>1$, since its modulus is
constant along directions orthogonal to $\v$.  But eq. (5) still makes
sense provided $\fh$ is sufficiently well--behaved.  (This is one of the
reasons we have assumed that $f$ is a test function.)  

\skp

A common method for the construction of $n$--dimensional wavelets
consists of  taking tensor products of one--dimensional wavelets.  However,
this means that not all directions in $\rn$ are treated equally, and
consequently the set of  wavelets does not transform ``naturally'' (in a sense
to be explained below)  under the {\sl affine  group  \/} $G$ of $\rn$, which
consists of all transformations of the form

$$
\x\mapsto g(A, \b)\x\equiv A\x+\b
\eqno(8)
 $$
with $A$ a non--singular $n\times n$ matrix and $\b\in\rn$.  Each such 
$ g(A, \b)$ defines a unitary operator on 
$L^2(\rn)$, given by

$$
 \left( U(A, \b)f \right)(\x)\equiv |A|^{-\h}f\left(A ^{-1} (\x-\b)  \right), 
\eqno(9)
 $$
where $|A|$ denotes the absolute value of the determinant of $A$.
The map \break
 $g(A, \b)\mapsto U(A, \b)$ forms a {\sl representation\/} 
of $G$ on  $L^2(\rn)$, meaning that it preserves the group structure of $G$
under compositions.  To see how $h_{\x, \v}$ transforms under $U$, note
that the unitarity of $U$ implies

$$\eqalign{
\l U(A, \b)\,h_{\x, \v}\, ,f\r & =\l h_{\x, \v} \, ,U(A, \b) ^{-1} f\r \cr
 & =\inr dt\,\, h(t)^*\, |A|^\h f\left( A(\x+t\v)+\b \right)\cr
 & =|A|^{\h} \l h_{A\x+\b, A\v} \, ,f\r.\cr}
 \eqno(10)
 $$
Hence 

$$
 U(A, \b)\,h_{\x, \v}=|A|^{\h} h_{A\x+\b, A\v},
\eqno(11)
 $$
which states that affine transformations take wavelets to wavelets. 
Thus, for example, translations, rotations and dilations merely translate,
rotate and dilate the labels $\{\x, \v\}$, while the factor $|A|^\h$  preserves
unitarity.  By contrast, tensor products of one--dimensional
wavelets are not  transformed into one another  by
rotations.

\sv2

\bf 2.2. A Reconstruction Formula \rm
\sv1

\noindent   A  reconstruction consists of a recovery of $f$ from $f_h$ or its
restriction to some   subset.  In the one--dimensional case, for
example, $f$ can be reconstructed using {\sl all\/}  of $\rr\times\rr_*$ or
(for certain choices of $h$) just a discrete subset [2, 6, 18, 21, 22].  For
general $n$, the choice of reconstructions becomes even richer since
various new possibilities arise.  For example, $h$ may have 
symmetries which imply that $f_h$ is determined by its values on some
lower--dimensional subsets of $\rn\times\rr^n_*$, making integration over
the whole space  unnecessary and, moreover, undesirable since it may lead
to a divergent integral.
 Furthermore,  $f$ may satisfy some partial differential equation
which implies that it is determined by its values on subsets of $\rn$.  For
example, if $\rn$ is space--time and $f$ represents a pressure wave or an
electromagnetic potential, it satisfies the wave equation away from sources,
hence is determined by initial data on a Cauchy surface in $\rn$, and it
becomes both unnecessary and undesirable to use all of $\rn\times \rr^n_*$ 
in the reconstruction (cf. Sections 3.2 and  4).

The reconstruction to be developed in this subsection is ``generic'' in that it
does not assume any particular forms for $h(t)$ or $f(\x)$. 
It uses all of $\rn\times \rr^n_*$, so it breaks down for certain
 choices of $h$  or $f$.   Again we emphasize that this is far from the only 
way to proceed;  other types of reconstruction will be discussed below and
elsewhere. The present reconstruction formula is  interesting in part because
it generalizes the one for the ordinary continuous wavelet transform $(n=1)$.

To reconstruct $f$, we look for a {\sl resolution of unity\/} in terms of the
vectors $h_{\x,\v}$.  This means we need a measure $d\mu (\x,\v)$ on
$\rn\times \rr^n_*$ such that 

$$
 \int_{\rn\times\rr^n_*} d\mu (\x,\v)\,|f_h(\x,\v)|^2=\int_\rn
d\x\,|f(\x)|^2\equiv \|f\|^2_{L^2}. 
\eqno(12)
 $$
(Such an identity is sometimes called a ``Plancherel formula.'')
For then the map $T{:}\,f\mapsto f_h$ is an isometry from $L^2(d\x)$ 
onto its range  $\ch\subset L^2(d\mu )$, and polarization gives

$$
 \langle\, g, T^*Tf\,\rangle_{L^2(d\x)}\equiv \langle\, Tg,Tf\,\rangle_\ch=
\langle\, g,f\,\rangle_{L^2(d\x)}. 
\eqno(13)
 $$
This shows that $f=T^*Tf=T^*f_h$ in $L^2(d\x)$,  which is the desired
reconstruction formula.  (Cf.  [16] for background on resolutions
of unity, generalized frames and related subjects.)  

 To obtain a resolution of unity, note that

$$
 f_h(\x,\v)=\left(\hh(\p\cdot\v)^* \,\fh(\p)\right)\check{\ }(\x),
\eqno(14)
 $$
where $\check{ }$ denotes the inverse Fourier transform, so by
Plancherel's theorem,

$$
 \int_\rn d\x\,|f_h(\x,\v)|^2=\int_\rn d\p\,|\hh(\p\cdot\v)|^2\,|\fh(\p)|^2.
\eqno(15)
 $$
We therefore need a measure $d\rho (\v)$ on $\rr^n_*$ such that 

$$
 H(\p)\equiv \int_{\rr^n_*} d\rho (\v)\,|\hh(\p\cdot\v)|^2\equiv 1 
\quad \hbox{for almost all }\p, 
\eqno(16)
 $$
since then $d\mu(\x, \v)=d\x\,d\rho (\v)$ has the desired property.
The solution is simple:  Every $\p\ne\o$ can be transformed to
${\bf q}\equiv (1, 0, \cdots, 0)$ by a  {\sl dilation and rotation\/} of $\rn$. 
That is, the orbit of ${\bf q}$ (in Fourier space) under dilations and rotations is
$\rr^n_*$.  Thus we choose $d\rho $ to be invariant under rotations and
dilations, which gives

$$
 d\rho (\v)=N|\v|^{-n} d\v,
\eqno(17)
 $$
where $N$ is a normalization constant, $|\v|$ is the Euclidean norm of
$\v$ and $d\v$ is Lebesgue measure in $\rn$.  Then for $\p\ne \o$,

$$\eqalign{
H(\p) & =H({\bf q})=N\int |\v|^{-n} d\v\,|\hh(v_1)|^2\cr
 & =N\inr dv_1\,|\hh(v_1)|^2\,
\int_{\rr^{n-1}} { dv_2 \cdots dv_n\over (v_1^2+\cdots v_n^2)^{n/2}}.
 \cr}
 \eqno(18)
 $$
Now a straightforward computation gives 

$$
 \int_{\rr^{n-1}} { dv_2 \cdots dv_n\over (v_1^2+\cdots v_n^2)^{n/2} }=
{ \pi ^{n/2}\over |v_1|\,\Gamma(n/2)}.
\eqno(19)
 $$
This shows that  the measure $d\mu (\x,\v)\equiv d\x \,d\rho (\v)$
gives a resolution of unity if and only if  

$$
 c_h\equiv \inr { d\xi \over |\xi |}\,|\hh (\xi )|^2<\infty,
\eqno(20)
 $$
which is precisely the {\sl admissibility condition \/} for the usual
(one--dimensional) wavelet transform [5].  (As mentioned above, 
admissibility implies that $\hh(0)=0$.)  If $h$ is admissible, the
normalization constant is given by

$$
N={ \Gamma(n/2)\over \pi  ^{n/2} \,c_h} 
\eqno(21)
 $$
and the reconstruction formula is

$$
 f(\x')=(T^*f_h)(\x')=N\int_{\rn\times\rr^n_*}  |\v|^{-n} d\x\,d\v\,
h_{\x,\v}(\x')\,\,f_h(\x,\v). 
\eqno(22)
 $$
The {\sl sense\/}  in which this formula holds depends on the
behavior of $f$.  The class of possible $f$'s, in turn, depends on the choice
of $h$.  Note that in spite of the factor $|\v|^n$ in the denominator, there is
no problem at $\v=\o$ since $ f_h(\x,\o)=\hh(0)^*\,f(\x)=0$
by the admissibility condition, and a similar analysis can be made for
small $|\v|$ by using the dilation property (eq.(3)).

\sv2

\bf 2.3. The $d$--Dimensional Windowed Radon Transform \rm
\sv1

\noindent Next, we allow the instrument to extend in $k\ge 0$ spatial
dimensions.  For example, $k=1$ for a wire antenna whereas $k=2$ for a dish
antenna. If $\rn$ is space, then $k\le n$;  if $\rn$ is space--time, then $k\le
n-1$.   When moving through space with a uniform velocity, the instrument 
sweeps out a $d$--dimensional surface in $\rn$, where $d=k+1$ if the
motion is transverse to its spatial extension and $d=k$ if it is not. If $k<n$,
then the set of non--transversal motions is ``non--generic'' (has measure
zero) and can thus be ignored; we therefore set $d=k+1$ in that case.  If
$k=n$, then necessarily $d=n$.  In either case, we represent the moving
instrument by a window function $h:\/\rd\to\cc$. 

The parameter $t\in\rr$ has thus been replaced by $\t\in\rd$.  The velocity
vector $\v$, which may be regarded as a linear map $t\mapsto \v t$ from
$\rr$ to $\rn$, is now replaced by a linear map $A:\/\rd\to\rn$,  which we
call a {\sl motion\/}  of the instrument in $\rn$. Denote the
set of all such maps by $L(\rd, \rn)$.  Later, when seeking reconstruction,
we shall need to restrict ourselves to subsets of $L(\rd, \rn)$ (`rigid'
motions); but this need not concern us presently.

\skp

\proclaim Definition 2.  The d--dimensional Windowed Radon Transform of
$f:\/ \rn\to\cc$ is the function $f_h:\/\rn\times L(\rd, \rn)\to\cc$
given by
$$
 f_h(\x, A)=\int_\rd d\t\,h(\t)^*\,f(\x+A\t).
\eqno(23)
 $$

Upon substituting the Fourier representation of $f$, a computation similar to
the above yields the expression

$$\eqalign{
f_h(\x, A) & =\int_\rd d\t\,\int_\rn d\p\,e^{2\pi i\p\cdot(\x+A\t)}
\, h(\t)^*\,\fh(\p)\cr
 & =\int_\rn d\p\,e^{2\pi i\p\cdot\x}\,\hh(A'\p)^*\,\fh(\p)\cr  
 & \equiv \langle\, \hh_{\x,A}\,  , \fh\,\rangle_{L^2(d\p)}=
\langle\,  h_{\x,A}\,  , f\,\rangle_{L^2(d\x)},
 \cr}
 \eqno(24)
 $$ 
where  $A':\/\rn\to\rd$ is the map dual to $A$. (For given bases in $\rd$
and $\rn$, $A$ is represented by an $n\times d$ matrix; then $A'$ is the
transposed $d\times n$ matrix.)  In the above equation we have set

$$
 \hh_{\x,A}(\p)=e^{-2\pi i \p\cdot\x}\,\hh(A'\p),
\eqno(25)
 $$
which gives the generalized wavelets

$$
 h_{\x,A}(\x')=\int_\rn d\p\,e^{2\pi i\p\cdot(\x'-\x)}\,\hh(A'\p).
\eqno(26)
 $$

Let us now attempt to reconstruct $f$ from $f_h$ by generalizing the
procedure in Section 2.2.    Eq. (15) now becomes

$$
 \int_\rn d\x\,|f_h(\x,A)|^2=\int_\rn d\p\,|\hh(A'\p)|^2\,|\fh(\p)|^2.
\eqno(27)
 $$
Again we need a measure $d\rho (A)$ which is invariant under dilations
and rotations of $\rn$. Now the {\sl largest\/}  set of maps $A$ which
can be considered consists of all those with rank $d$.  (Otherwise the
instrument sweeps out a surface of dimension lower than $d$.)  Let us
call this set $L_d(\rd, \rn)$.   Then a measure on $L_d(\rd, \rn)$ which
is invariant  with respect to rotations and dilations of $\rn$ 
has the form

$$
 d\tilde\rho (A)=|\det(A'A)|^{-n/2}\,dA,
\eqno(28)
 $$
where $dA$ is the Haar measure on $L(\rd, \rn)\approx\rr^{nd}$ as
an additive group.  However, no reconstruction is possible using the
measure $d\x\, d\tilde\rho (A)$ on $\rn\times L_d(\rd, \rn)$, because 
{\sl no admissible window exists\/} in general when $d>1$.  (This  can
be easily verified when $d=n=2$.)  Thus $L_d(\rd, \rn)$ is too large,
and we return to our imaginary measuring process for inspiration.  On
physical grounds, we are interested in {\sl rigid motions\/}  of the
instrument. A map corresponding to such a motion  must have the form
$A=vRJ$, where $J :\,\rd\to\rn$ is the canonical inclusion map, $R$ is a
rotation of $\rn$ ($RJ$ gives the orientation of the instrument as well as its
direction of motion), and $v> 0$ is the speed. (If $\rn$ is space--time, then
``rotations'' involving the time axis are actually Lorentz transformations!  For
the present, assume that $\rn$ is space, so $R$ is a true rotation.) We
therefore parametrize the set of permissible $A$'s by $(v, R)\in \rr^+\times
SO(n)\equiv G$, where  $SO(n)$ is the group of unimodular orthogonal
$n\times n$ matrices, which represent rotations in $\rn$.  This
parametrization is redundant because two rotations of $\rn$ which have the
same effect on the subspace $\rd$ give  the same motion.  A non--redundant
parametrization of rigid motions is given by $\rr^+\times
\left(SO(n)/SO(n-d)  \right)$.    However, we  use the redundant one
here for simplicity.  (We shall need the Haar measure on $SO(n)$.)
 Note that for $d=1$,  $J$ is represented by the vector $\q=(1, 0, \cdots, 0)$
and  the set of all  maps $A=vRJ$ as above coincides with the set $\rr^n_*$ of
non--zero velocities considered in Sections 2.1 and 2.2. 
A measure on $G$ which is invariant under  rotations
and dilations of $\rn$ (\ie under $G$ itself) has the form

$$
 d\rho (A)=Nv ^{-1} dv\,dR,
\eqno(29)
 $$
where $N$ is a normalization constant and $dR$ is the Haar measure on
$SO(n)$.  Thus for all $\p\ne\o$,

 $$
 H(\p)\equiv \int_G d\rho (A)\,|\hh(A'\p)|^2=H(\q).
\eqno(30)
 $$
 Now 

$$
 A'\q=vJ'R'\q=vJ'{\bf R}_1',
\eqno(31)
 $$
where ${\bf R}_1$ is the first row of $R$, which is a unit vector, and $J'$ is
the projection of $\rn$ onto $\rd$.  The admissibility condition therefore
 reads

$$
 N ^{-1} \equiv \int_0^\infty v ^{-1} dv \int_{SO(n)} dR\,|\hh(vJ'{\bf R_1'})|^2
<\infty.
\eqno(32)
 $$
For $d=1$, this reduces to eq. (20). If $h$ is admissible, we obtain the
reconstruction formula

$$
 f(\x')=\int_\rn  d\x\int_G d\rho (A)\,h_{\x,A}(\x')\,\,f_h(\x,A).
\eqno(33)
 $$

\sv4
\cl{\bf 3. Analytic--Signal Transforms}
\sv1

\bf 3.1. Analytic Signals in One Dimension \rm
\sv1

\noindent Suppose we are given a (possibly complex--valued)
one--dimensional signal \break
 $f: \,\rr\to\cc$.  For simplicity, assume that
$f$ is smooth with rapid decay. Consider the positive-- and negative--
frequency parts of $f$, defined by

$$\eqalign{
f^+(x) & \equiv \int_0^\infty dp \,e^{2\pi ipx}\,\fh(p )\cr
f^-(x) & \equiv \int_{-\infty}^0 dp \,e^{2\pi ipx}\,\fh(p ),
\cr}
 \eqno(34)
 $$
Then $f^+$ and $f^-$ extend analytically to the upper--half and lower--half
complex planes, respectively; \ie

$$\eqalign{
f^+(x+iy) & = \int_0^\infty dp\,e^{2\pi ip(x+iy)}\,\fh(p),\quad y>0\cr
f^-(x+iy) & = \int_{-\infty}^0 dp\,e^{2\pi ip(x+iy)}\,\fh(p),\quad y<0,
 \cr}
 \eqno(35)
 $$
since the factor $e^{-2\pi py}$ decays rapidly for $p\to\pm\infty$ in the
respective integrals.   $f^+(z)$ and $f^-(z)$ are just the (inverse) {\sl
Fourier--Laplace transforms\/}  of the restrictions of $\fh$ to the positive
and negative frequencies.
If $f$ is complex--valued, then $f^+$ and $f^-$ are independent
and the original signal  can be recovered from them as

$$
 f(x)=\lim_{y \downarrow 0}\left[ f^+(x+iy)+ f^-(x-iy ) \right].
\eqno(36)
 $$ 
If $f$ is real--valued, then 

$$
 \fh(p)=\fh(-p)^*.
\eqno(37)
 $$
In that case, $f^+$ and $f^-$ are related by reflection,
$$
 f^+(x+iy)=f^-(x-iy)^*, \quad y>0,
\eqno(38)
 $$
and

$$
 f(x)=2\lim_{y \downarrow 0} \Re f^+(x+iy )
=2\lim_{y \downarrow 0} \Re f^-(x-iy ).
\eqno(39)
 $$
When $f$ is real, the function $f^+(z)$ is known as the {\sl analytic signal\/} 
associated with $f(x)$. Such functions were first introduced and applied
extensively by Gabor  [8].   A complex--valued signal would have {\sl two\/} 
independent associated analytic signals $f^+$ and $f^-$.  What significance do
$f^\pm$ have?  For one thing, they are {\sl regularizations\/}  of $f$.  Eq.
(36) states that $f$ is jointly  a ``boundary--value'' of the pair $f^+$ and
$f^-$.  As such, $f$ may actually be quite singular while remaining the
boundary--value of analytic functions.  Also,  $2f^\pm$ provide a kind of 
 ``envelope'' description of  $f$ (cf. Born and Wolf [4],  Klauder and
Sudarshan [19]). 
 For example, if $f(x)=\cos x$, then $2f^\pm(x)=e^{\pm ix}$.

In order to extend the concept of analytic signals to more than one
dimension, let us first of all unify the definitions of $f^+$ and $f^-$ by
setting 

$$
 \ftil(x+iy)\equiv \int_{-\infty}^\infty dp\,\theta
(py)\,e^{2\pi ip(x+iy)}\,\fh(p)
 \eqno(40)
 $$
for {\sl arbitrary\/}  $x+iy\in\cc$, where $\theta $ is the 
unit step function,  defined by

$$
 \theta (u)=\cases{0,\quad &$u<0$\cr \cr \h,\quad &$u=0$\cr \cr1,\quad &
$u>0$.\cr}
 \eqno(41)
 $$
Then we have

$$\eqalign{
\ftil(x+iy)=\cases{f^+(x+iy),\quad &$y>0$\cr\cr \h f(x),\quad &$y=0$\cr\cr
 f^-(x+iy),\quad &$y<0.$\cr}
 \cr}
 \eqno(42)
 $$

 Although this unification of $f^+$ and $f^-$ may at first appear to
be somewhat artificial, it  turns out to be quite natural, as will now be seen. 
Note first of all that for any real $u$, we have 

$$
 \theta (u)\,e^{-2\pi u}={ 1\over 2\pi i}\inr{ d\tau \over \tau -i} \,
e^{2\pi i\tau u},
\eqno(43)
 $$
since the contour on the right--hand side  may be closed 
in the upper--half  plane when $u>0$ and in the lower--half plane when
$u<0$.   For $u=0$, the equation states that 

$$\eqalign{
\theta (0) & ={ 1\over 2\pi i} \int_{-\infty}^\infty {(\tau +i)\,d\tau 
 \over \tau ^2+1}\cr
 & ={1 \over 2\pi } \int_{-\infty}^\infty { d\tau \over \tau ^2+1}=\h,
 \cr}
 \eqno(44)
 $$
 in agreement with our definition, if we interpret the integral as 
the limit when $L\to\infty$ of the integral from $-L$ to $L$. Therefore
$$
 \theta(py)\,e^{2\pi ip(x+iy)}=\inds  e^{2\pi ip(x+\tau y)}.
\eqno(45)
 $$
If this is substituted into our expression for $\ftil(z)$ and the order of
integrations on $\tau $ and $p$ is exchanged, we obtain

$$
 \ftil(x+iy)=\inds f(x+\tau y)
\eqno(46)
 $$
for arbitrary $x+iy\in\cc$.   We have referred to the right--hand side   as the
{\sl Analytic--Signal transform \/}  of $f(x)$ [16, 17].  It bears a close
relation to the {\sl Hilbert transform,\/} which is defined by

$$
 Hf(x)={ 1\over \pi }{\rm PV}\int_{-\infty}^\infty
 {du \over u}\,f (x-u), 
\eqno(47)
 $$
where PV denotes the principal value of the integral.  Consider the complex
combination 

$$\eqalign{
 f (x)-iHf(x) & ={1\over \pi i}\int_{-\infty}^\infty du
\left[ \pi i\delta (u) +{\rm PV} {1 \over u}\right]\,f(x-u)\cr
 & ={1\over \pi i}\lim_{\epsilon \downarrow 0}\,\int_{-\infty}^\infty 
{ du\over u-i\epsilon }\,f (x-u)\cr
 & ={1\over \pi i}\lim_{\epsilon \downarrow 0}\,\int_{-\infty}^\infty 
{ d\tau \over {\tau -i}}\,f (x-\tau \epsilon )\cr
 & =2\lim_{\epsilon \downarrow 0}\,\ftil (x-i\epsilon ).
 \cr}
 \eqno(48)
 $$
Similarly, 

$$
  f (x)+iHf(x)=2\lim_{\epsilon \downarrow 0}\,\ftil (x+i\epsilon ).
\eqno(49)
 $$
Hence

$$
Hf(x)=i\lim_{\epsilon \downarrow 0}\,[\ftil(x-i\epsilon )-
\ftil(x+i\epsilon )],
\eqno(50)
 $$
which for real--valued $f$ reduces to 

$$
 Hf(x)=2\lim_{\epsilon \downarrow 0}\,\Im \ftil(x+i\epsilon )
=-2\lim_{\epsilon \downarrow 0}\,\Im \ftil(x-i\epsilon ).
\eqno(51)
 $$

\sv2

\bf 3.2.  Generalization to n Dimensions \rm
\sv1

\noindent We are now ready to generalize the idea of analytic signals to an
arbitrary number of dimensions.  Again we assume initially that $f(\x)$
belongs to the space of Schwartz test functions ${\cal S}(\rn)$, although this
assumption  proves to be unnecessary.

 \skp

\proclaim Definition 3. The  Analytic--Signal Transform (AST)
 of $f\in{\cal S}(\rn)$ is the function $\ftil:\cn\to\cc$  defined by

$$
 \ftil(\x+i\y)=\inds f(\x+\tau \y).
\eqno(52)
 $$

\skp 

\noindent The same argument as above shows that for $\z=\x+i\y\in\cn$,

$$\eqalign{
 \ftil(\z) & =\inp  \eyp  \,\fh(\p)\cr
 & =\int_{M_\y} d\p\, e^{2\pi i\p\cdot\z}\,\fh(\p),
\cr}
 \eqno(53)
 $$
where  $M_\y$ is the half--space

$$\eqalign{
M_\y\equiv \{\p\in \rn \st \p\cdot \y\ge 0\},\qquad \y\ne\o.
 \cr}
 \eqno(54)
 $$

We shall refer to the right--hand side  of eq. (53) as the (inverse)  {\sl
Fourier--Laplace transform\/}    of $\fh$ in $M_\y$.
 The integral converges absolutely whenever
$\fh \in L^1(\rn)$, since  $|e^{2\pi i\p\cdot\z}|\le  1$ on $M_\y$,
defining $\ftil$ as a {\sl function\/}  on $\cn$, although not an analytic
one in general (see below).  This shows that
 $\ftil(\z)$ can actually be defined for some distributions $f$, not only for
test functions.  

\skp

\noindent {\sl Note:\/} In spite of the appearance of expressions such as 
$\p\cdot \y$, we have not assumed any particular metric structure in
$\rn$.  The Fourier transform naturally takes functions on $\rn$
(considered as an abelian group) to functions on the {\sl dual\/}  space
$\rr_n\equiv {(\rn)}^*$ of linear functionals, and $\p\cdot\y$ merely
denotes the value $\p(\y)$.  (See Rudin [25].) This remark becomes
especially important when considering time--dependent signals, so that
$\rn$ is space--time, for then the natural structure on $\rn$ is a Lorentzian
metric rather than a Euclidean metric (cf.  [16], Section 1.1.)
 \skp

For $n=1$, $\ftil(z)$ was analytic in the upper-- and lower--half planes.  In
more than one dimension, $\ftil(\z)$ need not be analytic, even though, for
brevity, we still write it as a function of $\z$ rather than $\z$ and its
complex conjugate $\z^*$.  However, $\ftil(\z)$ does in general possess a {\sl
partial\/} analyticity which reduces to the above when $n=1$.  Consider the
partial derivative of $\ftil(\z)$ with respect to $z^*_k\equiv x_k-iy_k $,
defined by

$$\eqalign{
2\bar \partial_k \ftil\equiv  2{\partial \ftil\over{\partial z^*_k}}\equiv 
{\partial\ftil\over{\partial x_k}}+i{\partial\ftil\over{\partial y_k}}.  
 \cr}
 \eqno(55)
 $$
Then $\ftil$ is analytic at $\z$ if and only if $\bar\partial_k \ftil=0$ for all
$k$.  But using our definition of $\ftil(\z)$, we find that 

$$\eqalign{
2\bar \partial_k \ftil(\z)={ 1\over 2\pi }\inr d\tau \,
{\partial f\over{\partial x_k}}(\x+\tau \y).
 \cr}
 \eqno(56)
 $$
 It follows that  the complex  $\bar \partial$--derivative  in the direction
of $\y$ vanishes,   i.e.

$$\eqalign{
4\pi \y\cdot\bar\partial\ftil(\z)\equiv 
4\pi \sum_k y_k \,\bar \partial_k \ftil(\z) &
=\inr d\tau \, \sum_k y_k{\partial f\over{\partial x_k }}(\x+\tau \y)\cr
 & =\inr d\tau \,{\partial \over{\partial \tau }}f(\x+\tau \y)=0,
 \cr}
 \eqno(57)
 $$
if $f$ decays for large $|\x|$ (for example, if $f$ is a test function, as we
have assumed).  Equivalently, using

$$\eqalign{
2\bar\partial_k \,\left[ \eyp\right]
 & =2\bar\partial_k\,\left[ \theta (\p\cdot \y)\right]\,e^{2\pi i\p\cdot\z}\cr 
 & =i{\partial \theta (\p\cdot \y)\over{\partial y_k }}\,\,e^{2\pi i\p\cdot
\z}\cr 
 & =ip_k\,\delta (\p\cdot \y)\,e^{2\pi i\p\cdot \z}\cr
 & =ip_k \,\delta (\p\cdot \y)\,e^{2\pi i\p\cdot \x},
 \cr}
 \eqno(58)
 $$
we have for $\y\ne \o$

$$\eqalign{
2\y\cdot\bar\partial\ftil(\z)= i\inp (\p\cdot \y)\,\delta (\p\cdot
\y) \,e^{2\pi i\p\cdot \x}\,\fh(\p)=0.
\cr}
 \eqno(59)
 $$
Thus  $\ftil(\z)$ {\sl  is analytic in the direction\/}  $\y$.
In the one--dimensional case, this reduces to

$$
{\partial \ftil(z)\over{\partial z^*}}=0\quad \forall y\ne 0,
\eqno(60)
 $$
which states that $\ftil(z)$ is analytic in the upper-- and lower--half planes. 
In one dimension, there are only {\sl two\/}  imaginary
directions, whereas in $n$ dimensions, every $\y\ne \o$
defines an imaginary direction. 
\skp

The multivariate AST is related to the {\sl Hilbert
transform in the direction\/}  $\y$  (cf. [26], p. 49),  defined as

$$\eqalign{
 H_\y f(\x)={ 1\over \pi }{\rm PV}\inr
 {du \over u}\,f(\x-u\y), \quad \x,\y\in\rn,\ \y\ne \o.
 \cr}
 \eqno(61)
 $$
(Usually, it is assumed that $\y$ is a unit vector; we do not make this
assumption.)   Namely, an argument similar to the above shows that 

$$
f(\x)\pm iH_\y f(\x)=2\lim_{\epsilon \downarrow 0}\,\ftil (\x\pm i\epsilon
\y), 
\eqno(62)
 $$
hence

$$
H_\y f(\x)=i\lim_{\epsilon \downarrow 0}\,[\ftil(\x-i\epsilon \y)-
\ftil(\x+i\epsilon \y)].
\eqno(63)
 $$
For $n=1$ and $y>0$, this reduces to the previous relation with the
ordinary Hilbert transform. 

 As in the one--dimensional case,  $f(\x)$ is the 
boundary--value of $\ftil(\z)$ in the sense that 

$$
f(\x)=\lim_{\epsilon \to 0}\,[\ftil(\x+i\epsilon \y)+
\ftil(\x-i\epsilon \y)].
\eqno(64)
 $$
For real--valued $f$, these equations become

$$\eqalign{
f(\x)=2\lim_{\epsilon \to 0}\,\Re\, \ftil(\x+i\epsilon \y)\cr
H_\y f(x)=2\lim_{\epsilon \downarrow 0}\,\Im\,\ftil(\x+i\epsilon \y).
 \cr}
 \eqno(65)
 $$

\skp

\noindent Two unresolved yet fundamental questions are:
\skp

\item{(a)}   For what classes of `functions' (possibly distributions) can the
AST be defined, apart from ${\cal S}(\rn)$; \ie what is the {\sl domain\/}  of
the AST? 

\item{(b)}  Given a vector space $\ch$ of `functions'  on $\rn$
for which the AST is defined, what is the {\sl range\/}  of the AST on $\ch$? 
That is, given a function $F$ on $\cn$, how can we tell whether $F$ is the
transform of some $f\in\ch$? 

\skp

A necessary, though probably not sufficient,  condition for $F=\ftil$ is that
$F$ satisfy the directional Cauchy--Riemann equation $\y\cdot{\bf
\bar\partial}\,F(\x, \y)=0$.  Complete answers to the above questions can be
given in some specific cases:  When $f$ is a solution of the Klein--Gordon
equation, then $\ftil$ must satisfy a certain consistency condition (the
reproducing property of the associated wavelets).  This condition, when
satisfied by $F$, also guarantees that $F=\ftil$ for some $f$ (cf. [16],
Chapters 1 and 4).  A similar result will be obtained in Section 4 for
solutions of the wave equation in two space--time dimensions.  The
comments below apply to the general case and are, consequently, informal.

 The most direct way to find if  $F$ is the AST of some
$f$ is to {\sl construct\/} $f$ from $F$ and then check that $\ftil=F$.  The
first part has already been done formally, since $f$ has been shown to be
the boundary--value of $\ftil$.   Here we suggest an alternative method
which can be used to construct $\fh(\p)$ instead of $f(\x)$.   Assume that
the Fourier transform is defined on  $\ch$.  If $F(\x, \y)=\ftil(\x+i\y)$ for
some $f(\x)\in\ch$, then the $2n$--dimensional Fourier transform of $F$ is
seen (formally) to be

$$\eqalign{
 \hat F(\p, \q)&\equiv \int_{\rr^{2n}} d\x\,d\y\,e^{-2\pi i(\p\cdot
\x+\q\cdot\y)}\, F(\x, \y)\cr
&=\fh(\p)\,\int_\rn d\y\,\theta (\p\cdot\y)\,e^{-2\pi i(\q-i\p)\cdot\y}\cr
&\equiv \fh(\p) \,\tilde \delta (\q-i\p),
 \cr}
 \eqno(66)
  $$
where $\tilde \delta $ is the AST, {\sl in Fourier space,\/}  of the Dirac 
measure $\delta (\q)$.   (This suggests that the AST, like the Fourier
transform, exhibits some symmetry between space and Fourier space.) 
$\tilde \delta $ can be shown to be invariant under {\sl real\/}  rotations
($\q-i\p\to R(\q-i\p)$, with $R\in SO(n)$) and to transform under dilations
as 

$$
 \tilde \delta (\lambda (\q-i\p))=\lambda ^{-n}\,\tilde \delta (\q-i\p),
\qquad \lambda \ne 0.
\eqno(67)
 $$
Given $\p\ne\o$, let $R$ be a rotation such that $\p=|\p|\,R\,\e$, where
$\e=(1, 0, \cdots,0)$, and let $\k\equiv |\p|^{-1} R ^{-1} \q$, so that 
$\q-i\p=|\p|\,R\,(\k-i\e)$.  Then

$$\eqalign{
\tilde \delta (\q-i\p)&=|\p|^{-n}\,\tilde \delta (\k-i\e)\cr
&=|\p|^{-n}\,\int_\rn d\y\,\theta (y_1)\,e^{-2\pi i(\k-i\e)\cdot \y}\cr
&=|\p|^{-n}\,\int_0^\infty dy_1\,e^{-2\pi(1+ik_1)y_1}
\int_{\rr^{n-1}}dy_2\,\cdots  dy_n\,e^{-2\pi
i(k_2y_2+\cdots+k_ny_n)}\cr &={\delta (k_2, \cdots, k_n) \over 2\pi 
|\p|^n(1+ik_1)}.
 \cr}
 \eqno(68)
  $$
Let $P:\rn\to\rr$  and $Q:\rn\to\rr^{n-1}$ denote the projections $P\k=k_1$
and $Q\k=(k_2,\cdots, k_n)$.  Then the numerator in the last expression is

$$
 \delta(Q\k)=\delta(Q|\p|^{-1} R ^{-1} \q)|=|\p|^{n-1}\,\delta (QR ^{-1} \q),
\eqno(69)
 $$
hence

$$
 \tilde \delta (\q-i\p)={ \delta (QR ^{-1} \q)\over 2\pi (|\p|+iPR^{-1} \q)},
\qquad \p\ne\o.
\eqno(70)
 $$
Together with eq. (66), this gives an explicit formal condition for $F$ to be
the AST of $f$ and, if so, to determine $\fh(\p)$.  When $n=1$, $\tilde
\delta$ takes the simple form

$$
 \tilde \delta (q-ip)={ {\rm sgn}(p)\over 2\pi (p+iq)}\,  , \qquad p\ne 0.
\eqno(71)
 $$

\sv2

\bf 3.3. Some  Applications of the AST \rm

\sv1

\noindent The AST  is an example of a  Windowed
X--Ray transform,  introduced in Section 2.1, with  the
window  function

$$
 h(\tau )^*={1 \over 2\pi i(\tau -i)}.
\eqno(72)
 $$
(This window function  is not ``admissible'' in the  sense of Section 2.2,
hence the  reconstruction formula developed there  fails. 
However, that reconstruction was based on some assumptions which may
not be appropriate in general;  see the comments below Eq. (88).)  We now
give two examples of the usefulness of the AST.  An extensive use of this
transform will also be made in Section 4.
 
\VE

\noindent {\sl Example 1:  Hardy spaces}
\skp

\noindent Suppose that $\fh(\p)$
vanishes outside of some closed convex cone $V_+$.  The cone  $V_+'$ {\sl
dual\/} to $V_+$ is defined as 
$$
 V_+'=\{\y\in\rn\st\p\cdot\y>0\ \forall \p\in V_+ \},
\eqno(73)
 $$
and it is clearly an open convex cone. Note that for $\y\in V_+'$, $\theta
(\p\cdot\y)\equiv 1$  on the support of $\fh$ (except at $\p=\o$, which
has measure 0), hence if $f\in L^2(\rn)$ and $\y\in V_+'$, then

$$
 \ftil(\x+i\y)=\int_{V_+} d\p\, e^{2\pi i\p\cdot (\x+i\y)}\,\fh(\p)
\eqno(74)
 $$
 and it follows (Stein and Weiss [27]) that $\ftil(\z)$
is analytic in the {\sl tube domain\/} 

$$
 \tb^+\equiv \{\x+i\y\in\cn\st\y\in V_+'\}.
\eqno(75)
 $$ 
The set $H^2\equiv \{\ftil\st  \fh\in L^2(V_+)\}$ is known as a {\sl Hardy
space.\/} Note also that $\ftil$ vanishes in the tube 

$$
 \tb^-\equiv \{\x+i\y\in\cn\st -\y\in V'_+\},
\eqno(76)
 $$
since there $\p\cdot\y<0$ for all $\o\ne\p\in V_+$.  Eq. (74) gives

$$
 f(\x)=\lim_{\epsilon \downarrow 0}\ftil(\x+i\epsilon \y),\qquad \y\in
V'_+\,  , 
\eqno(77)
 $$
which states that $f$ is a  boundary--value of $\ftil$.
Since $\ftil(\z)$ is analytic, it may be regarded as  a {\sl regularization\/}  of
$f(\x)$ (the latter, being merely square--integrable, is a distribution).  Eq.
(77) can be viewed as a ``reconstruction'' of $f$ from $\ftil$, albeit a
somewhat trivial one. 
\skp

\noindent{\sl  Example 2: The Klein--Gordon Equation}
\skp

\noindent An important application of the AST is to signals
that satisfy some partial differential equations.  (In fact, it was in this
context that the transform originated.)  Suppose that $f$ is a solution of
 the {\sl Klein--Gordon equation \/} in $\rn$,

 $$
\del f+m^2c^4f=0,
\eqno(78)
 $$
where

$$
 \del\equiv {\partial^2  \over{\partial x_1^2}}-c^2{\partial^2 
\over{\partial x_2^2}} -\cdots -c^2{\partial^2  \over{\partial x_n^2}}
\eqno(79)
 $$
is the D'Alembertian or wave operator for waves with propagation speed $c$.
Here $\rn$ is interpreted as {\sl space--time,\/}  with $x_1$ the
time coordinate and $(x_2, \cdots, x_n)$ the space coordinates,    and $m>0$
is a mass parameter.  This equation describes free relativistic particles in
quantum mechanics.  The limit $m\to 0$ gives the wave equation, which
will be discussed below.   Define the {\sl solid light cone\/} in Fourier space
by

$$
 V=\{\p\in\rn\st \p^2\equiv p_1^2-c^2p_2^2-\cdots c^2p_n^2\ge 0\}.
\eqno(80)
 $$
(Note that we are now using a Lorentz metric!)
$V$ is the union of the {\sl forward\/}  and {\sl backward\/}  light cones
$V_+$ and $V_-$, where $p_1\ge 0$ and $p_1\le 0$, respectively.  
Note that  $V_\pm$ are  convex but $V$ is not. The fact
that $f$ satisfies the Klein--Gordon equation  means that its Fourier
transform $\fh(\p)$ is supported on the   double mass hyperboloid

$$
 \om=\{\p\in\rn\st \p^2=m^2c^4\}=\omp\cup\omm,
\eqno(81)
 $$
where $\Omega _m^\pm\subset V_\pm$.  Thus $\fh=\fh^+ +\fh^-$,
where $\fh^\pm$ are distributions supported on $\Omega _m^\pm$.   Since
$\Omega _m^\pm\subset V_\pm$,  an argument similar to that used for 
Hardy spaces shows that the corresponding solutions $f^\pm(\x)$ have AST's
$\ftil^\pm(\z)$ which are analytic in $\tb^\pm$ and vanish in $\tb^\mp$,
where 

$$
 \tb^\pm=\{\x+i\y\in\cn\st \y\in V_\pm'\}
\eqno(82)
 $$
and  $V_\pm'$ are the cones dual to $V_\pm$, which can be seen to be
$$
 V_\pm'=\{\y\in\rn\st \y^2\equiv c^2y_1^2-y_2^2-\cdots -y_n^2> 0, 
\quad \pm y_1>0\}.
\eqno(83)
 $$
Note that while $V$ is a cone in {\sl Fourier space\/} (i.e., $p_1$ is a
frequency and $p_2, \cdots p_n$ are wave numbers per unit length),
$V'\equiv V_+'\cup V_-'$ is a cone in space--time.  Technically, these two
spaces are  dual  and should not be identified with one another.  

The AST of $f$, given by $\ftil(\z)=\ftil^+(\z)+\ftil^-(\z)$, is therefore
analytic in the {\sl double tube\/}  $\tb\equiv \tb^+\cup\tb^-$, with $\tb^+$
and $\tb^-$ containing only the positive-- and negative--frequency parts of
$f$, respectively.  This ``polarization'' of frequencies is important because it
makes it possible to reconstruct the solution $f$ from $\ftil$ without
approaching the singular boundary $\rn\ (\y\to\o)$.
Eq. (74) shows that $\ftil^\pm$ have the form

$$
 \ftil^\pm(\z)=\int_{\Omega _m^\pm} d\tilde p\,\,
\hat e_\z^\pm(\p)^*\,a^\pm(\p), \qquad \z\in\tb^\pm,
\eqno(84)
 $$
where 

$$
 d\tilde p={ dp_2\,dp_3\cdots dp_n\over 2|p_1|}\equiv 
{ dp_2\,dp_3\cdots dp_n\over 2c\sqrt{m^2c^2+p_2^2+\cdots+p_n^2}}
\eqno(85)
 $$
 is the induced measure on $\om$ and 

$$
 \hat e_\z^\pm(\p)^*\equiv e^{2\pi i\p\cdot\z},\qquad \z\in\tb^\pm,\
\p\in \Omega _m^\pm.
\eqno(86)
 $$
The corresponding expression $e_\z^\pm$ in the space--time domain,
defined by

$$
 e_\z^\pm(\x')^*=\int_{\Omega ^\pm_m} d\tilde p\,\,e^{2\pi
i\p\cdot(\z-\x')}, 
\eqno(87)
 $$
is a solution of the Klein--Gordon equation  which can be shown to be a
{\sl coherent wave--packet\/}  whose parameters $\z=\x+i\y$ have a direct
geometric interpretation: $\x$ is a point in space--time about which
$e^\pm_\z$ is  ``focused''  (i.e., $e^\pm_\z(\x')$ converges toward the 
point $(x_2, \cdots , x_n)$ in space for times $x_1'<x_1$ and diverges away
from it for times $x_1'>x_1$), and $\y$ is a set of homogeneous coordinates
for the {\sl average  velocity\/}  at  which $e^\pm_\z$ is traveling. 
Furthermore, the invariant  $\lambda >0$ defined by $\lambda
^2\equiv \y^2$ can be interpreted as a {\sl scale parameter\/}  which,
roughly speaking, measures the spread (resolution) of the wave packet at
the instant of its maximal focus ($x_1'=x_1$) in its rest--frame (the
coordinate system in which $y_2=y_3=\cdots =y_n=0$).  For small $\lambda
$, $e^\pm_\z(\x')$ is localized near $x_k'=x_k\,\, (k=2, 3, \cdots, n)$ at time
$x_1'=x_1$, whereas for large $\lambda $,  it is spread out in space.  (In
Fourier space, on the other hand, it is $\lambda  ^{-1} $ which measures the
spread.)   The positive-- and negative--frequency packets $e_\z^+$ and
$e_\z^-$  are interpreted in quantum theory as particles and antiparticles,
respectively.  (This agrees with the usual observation that antiparticles ``go
backward in time.''  Cf.  [16], Chapter 5.)

\skp

Since the window function $h(\tau )$ used in the AST has Fourier transform
$\hat h(\xi )=\theta (\xi )\,e^{-2\pi \xi }$, it is not
admissible in the general sense developed  in Section 2.2; \ie we have

$$
 \inr { d\xi \over |\xi |}\,|\hat h(\xi )|^2=\infty.
\eqno(88)
 $$
However, the rules of the game have changed.
Eq. (88)  was associated with a reconstruction formula which represents
$f$ as an integral of generalized wavelets  parametrized by {\sl all\/}  of
$\rn\times\rr^n_*$,  \ie  all $\x+i\y$ with $\y\ne\o$.  This was
acceptable when considering general functions $f(\x)$ in $L^2(\rn)$,
since then we could define a representation of the affine group on such
functions.  But now we are dealing with a Hilbert space $\cal H$ of
solutions of the Klein--Gordon equation,

$$
 f(\x)=\int_{\Omega _m}d\tilde p\,e^{2\pi i\p\cdot\x}\,a(\p),
\eqno(89)
 $$
with the {\sl Sobolev\/}  norm 

$$
 \|f\|^2\equiv \int_{\Omega _m}d\tilde p\,|a(\p)|^2
\eqno(90)
 $$
rather the $L^2$ norm used in Section 2.2.
General affine transformations no longer map solutions to solutions,
i.e. they no longer define operators on $\cal H$.  The mass $m$ spoils the
invariance of the equation under dilations. Only the subgroup $\cp$ of
translations together with Lorentz transformations (i.e., linear maps
$\y\mapsto A\y$  which preserve the Lorentz norm $\y^2$) maps solutions
to solutions.  $\cp$ is called the inhomogeneous Lorentz or {\sl Poincar\'e\/} 
group.

 Recall that the measure used in the reconstruction formula of Section 2.2
was chosen to be invariant under dilations and rotations.  Since dilations no
longer define operators on $\cal H$, this measure is no longer appropriate. 
Rather, we now expect to reconstruct $f$ by integrating in $\y$ over the
double hyperboloid

$$
 \Omega _\lambda =\{\y\in V'\st \y^2=\lambda ^2\}=\Omega _\lambda
^+\cup\Omega _\lambda ^-
 \eqno(91)
 $$
for an arbitrary fixed $\lambda >0$.  Furthermore, we do not expect to
integrate over all $\x\in\rn$, since a solution is determined by its data on
any {\sl Cauchy surface\/}  $S\subset \rn$.  For simplicity, take $S$ to be
the hyperplane  $x_1=t$ for fixed $t\in\rr$, though any 
Cauchy surface (spacelike $(n-1)$--dimensional submanifold of $\rn$) will
 do  ([16], Section 4.5).  Thus consider the $(2n-2)$--dimensional
submanifold

$$
 \sigma =\{\x+i\y\in\tb\st x_1=t, \ \y\in\Omega _\lambda \}=\sigma
_+\cup \sigma _-,
 \eqno(92)
 $$
where $\y\in\Omega _\lambda ^\pm$ in $\sigma _\pm$.
$\sigma $  parametrizes all possible locations and velocities of a
classical particle at the fixed time $t$, i.e. it is a {\sl phase space.\/} 
 A reconstruction formula has been obtained in the form

$$
 f(\x')=\int_\sigma  d\mu (\z)\, e_\z(\x')\,\ftil(\z),
\eqno(93)
 $$
where $e_\z\equiv e_\z^\pm$ on $\sigma _\pm$,
 $\sigma_\pm $ is parametrized by $(x_2, \cdots x_n, y_2, \cdots y_n)$, and

$$
 d\mu (\z)=A(\lambda , m) ^{-1} dx_2\cdots dx_n dy_2\cdots dy_n.
\eqno(94)
 $$
 $A(\lambda , m)$ is a certain constant related to the admissibility of the
 wavelets $e_\z$ with respect to the measure $dx_2\cdots dy_n$.  
Note that this differs from the usual construction of a solution from its
initial data, which uses the values of both $f$ and $\partial f/\partial x_1$
on $S$.  The intuitive explanation is that the dependence of $\ftil(\x+i\y)$
on $\y\in \Omega _\lambda $, for fixed $\x\in S$, gives the equivalent
''velocity'' information.  The independence of the reconstruction from the
choice of Cauchy surface is due to a conservation law satisfied by solutions
(cf. [16]). 
\skp

 The above reconstruction formula   bears a close
resemblance to the standard representation of a function in terms of
wavelets, for the following reason:  In the hyperbolic geometry of spacetime,
a moving object undergoes a {\sl Lorentz contraction,\/}  i.e. it shrinks in its
direction of motion.  Since $\y$ represents a velocity, eq. (93) expresses $f$
as a linear combination of ``wavelets'' centered about all possible points in
space (at the given  time $t$) and in various states of compression.  However, 
the analogy is incomplete since the $e_\z$'s can only contract and not dilate. 
(That is, they have a minimum width in their rest frames,  determined by
the choice of $\lambda $.) Their contraction is due to Lorentz transformations
rather than ordinary dilations of the form $\x\mapsto a\x,\ a\ne 0$.
As noted earlier, the  Klein--Gordon equation is not invariant under such
dilations,  due to the presence of $m>0$.
On the other hand,  the wave equation $(m\to 0$) {\sl is\/}  invariant under 
dilations, hence the analogy with wavelets can be expected to be closer. 
This is the subject of our next  section.

\sv4

\cl{\bf 4.  Wavelets and The Wave Equation}
\sv1
\bf 4.1. Introduction \rm
\sv1

\noindent As explained in Section 1, a representation of solutions of the
wave equation similar to that given for the Klein--Gordon equation in 
(93) should be of some interest in the  analysis of naturally occurring
signals, since it would automatically display much of their informational
contents.  Unfortunately, the  reconstruction
formula in  (93) fails when $m\to 0$  because $A(\lambda , m)$ diverges
in that limit.  (The wavelet representation  is no longer square--integrable
in that limit.) This is probably related to the fact, well--known in quantum
mechanics, that the Klein--Gordon equation with $m>0$ has a very different
group--theoretical structure from the wave equation.  The symmetry group
of the Klein--Gordon equation is the Poincar\'e group $\cp$, while the
symmetry group of the wave equation  is the {\sl conformal group\/} $\cal
C$, which contains the Poincar\'e group as well as dilations and uniform
accelerations.   The fundamental difference between massive particles (such
as electrons) and massless particles (such as photons) is that the former can
be at rest while the latter necessarily travel at the speed of light.   It may
well be that once the conformal group is taken into account, an appropriate
reconstruction formula can be found.  In this section we confirm this
hypothesis in two--dimensional space--time.  

\sv2

\bf 4.2. Symmetries of the Wave and Dirac Equations in $\rr^2$ \rm
\sv1

\noindent In this subsection, we examine some group--theoretical aspects
of solutions of the wave equation in $\rr^2$. To simplify the notation, we
choose the units of length and time such that the propagation speed $c=1$. 
The wave  equation then reads

$$\eqalign{
0&=-\del f(x, t)\equiv \left( \partial_x^2-\partial_t^2 \right) \,f(x,t)\cr
 &=(\partial_x+\partial_t)(\partial_x-\partial_t)\,f(x,t),
 \cr}
 \eqno (95)
$$
and  we consider solutions which are possibly complex--valued.  
 In terms of the {\sl light--cone coordinates\/} 

$$
 u=x+t,\qquad v=x-t  ,
\eqno(96)
 $$
the equation becomes $\partial_u\partial_vf=0$,  hence the general
solution has the form first given by D'Alembert,

$$
 f(x, t)=f_+(u)+f_-(v).
\eqno(97)
 $$
  $f_+(t+x)$ is a {\sl left--moving wave\/} since it is constant on the
characteristics $x=x_0-t$, and similarly $f_-(t-x)$ is a {\sl right--moving
wave.\/}   Later we shall find an appropriate family of Hilbert spaces $\hs$
$(s\ge 0)$ to which $f_\pm$ will be required to belong, so we now write
$f_\pm\in\hs$ without being specific.  Note that we can let $f_\pm\to
f_\pm\pm c$, where $c$ is a constant, without affecting $f$.  This ambiguity
will not be a problem since $\hs$ contains no non--zero constants.  Hence
the solutions are in one--to--one correspondence with the elements of the
orthogonal sum $\cd_s=\hs\oplus\hs$, whose elements can be written
in the vector form

$$
\psi = \left(\matrix{f_+ \cr f_-	\cr}\right).
 \eqno(98)
 $$
The wave equation can be restated as $\partial_vf_+=\partial_uf_-=0$, or

$$\eqalign{
\left( \matrix{0 &\partial_t+\partial_x\cr \partial_t-\partial_x&0
		\cr}
  \right)\psi 
\equiv \left( \gamma _t\partial_t+\gamma _x\partial_x \right)\,\psi 
\equiv \dir \psi =0,
 \cr}
 \eqno(99)
  $$
where

$$
 \gamma _t=\left(\matrix{0&1\cr 1&0
		\cr}
 \right),\qquad  \gamma _x=\left(\matrix{0&1\cr -1&0
		\cr}
 \right).
\eqno(100)
 $$
Eq. (99) is known in quantum mechanics as the (two--dimensional, massless)
{\sl Dirac equation;\/}   $\psi $ is a {\sl  spinor,\/} and
$\gamma _t, \gamma _x$ are {\sl Dirac matrices.  \/} 
The Dirac equation may be viewed as a
particular system of  first--order equations equivalent to the wave
equation which is, moreover, especially suited to the symmetries of the
latter.  The Dirac operator $\dir$ is a ``square root'' of the wave operator
in the sense that $\dir^2=-\del I$, where $I$ is the $2\times 2$ identity
matrix.  (This is related to the Clifford algebra associated with the Lorentz
metric.)  In more than one space dimension, there is no such simple relation
between scalar--valued solutions $f$ and spinor--valued solutions $\psi$.

We denote by $\cd_{s+}$ and $\cd_{s-}$ the subspaces of $\cd_s$ with
vanishing second and first components, respectively.  Thus
$\cd_{s\pm}\approx\hs$.

 \skp

A {\sl symmetry\/}  of the wave equation is a transformation which maps
solutions to solutions.  As symmetries can be composed and inverted,
they form a group.   We shall be particularly interested in ``geometric''
symmetries, which are induced from mappings on  the underlying
space--time that  respect the wave equation. Some obvious
ones are:

\skp

\noindent{\sl  Translations:\/} For each $(x_0, t_0)\in\rr^2$, the map
$(x, t)\to(x+x_0, t+t_0)$ induces a symmetry transformation $f(x, t)\to
f(x-x_0, t-t_0)$.  This maps the right-- and left--moving waves by 

$$
 T(u_0,v_0):\ f_+(u)\to f_+(u-u_0),\qquad f_-(v)\to f_-(v-v_0),
\eqno(101)
 $$
where $u_0=x_0+t_0$ and $v_0=x_0-t_0$.  Hence  translations
can be made to act independently on the left and right parts of solutions.

\skp

\noindent{\sl  Lorentz transformations:\/}  For any real $\theta $, the
map 
$$
 x\to x\cosh \theta +t\sinh \theta ,\qquad 
 t\to x\sinh \theta +t\cosh \theta
\eqno(102)
 $$
preserves the Lorentz metric  $x^2-t^2=uv$.  (It is a
``rotation'' by the imaginary angle $-i\theta $ and represents the
space--time coordinates as measured by an observer moving with
velocity $-\tanh \theta $.)   This map has a particularly simple form in
terms of the light--cone coordinates: 
$$
 u\to e^\theta \,u\equiv \lambda u,
\qquad v\to e^{-\theta}\,v\equiv  \lambda^{-1}  v.
\eqno(103)
 $$
Since this leaves the wave operator $\del=-4\partial_u\partial_v$ 
invariant, it induces a symmetry  on solutions.  The simplest such
transformation acts on $f_\pm$ by $ f_+(u)\to f_+(\lambda^{-1}  u),\ 
f_-(v)\to f_-(\lambda  v)$. We shall need a more general induced
transformation, given by

$$
L(\lambda ):\  f_+(u)\to S_+(\lambda )\,f_+(\lambda^{-1}  u),
\qquad f_-(v)\to S_-(\lambda )\, f_-(\lambda  v), 
\eqno(104)
 $$
where $S_\pm :\,\rr^+\to\cc_*\equiv \cc\backslash \{0\}$ (called
``multipliers'') must satisfy

  $$S_\pm(\lambda  ^{-1} )=S_\pm(\lambda )^{-1},\qquad 
S_\pm(\lambda\lambda ')=S_\pm(\lambda )\,S_\pm(\lambda ')
\eqno(105)
 $$
in order that Lorentz transformations form a group.  (This means that
$S_\pm$ are group homomorphisms.)
Continuity in $\lambda >0$ then implies that $S_+(\lambda )=\lambda ^{-j}$
and $S_-(\lambda )=\lambda ^{-j'}$ for some $j, j'\in\cc$.  
\skp

The three--dimensional symmetry group of all maps $T(u_0,
v_0)\,L(\lambda )$ is the {\sl restricted Poincar\'e group\/} $\cp_0$ in
$1+1$ space--time dimensions.  Note that $\cp_0$ leaves invariant the
subspaces $\cd_{s\pm}$ of right-- and left--moving waves.  Since we shall be
interested in {\sl irreducible\/}  representations of the symmetry group
which characterize the complete wave equation, it is desirable to include a
symmetry which mixes these two subspaces.

\skp 

\noindent{\sl Space reflection:\/} 
The map $(x,t)\to(-x,t)$ is a discrete symmetry of the wave equation,
corresponding to  $(u, v)\to (-v,-u)$.  We take the induced mapping on
solutions to be 

$$
 P:\ f_+(u)\to f_-(-u),\qquad f_-(v)\to  f_+(-v).
\eqno(106)
 $$
 Thus, $P$ interchanges right and left waves, as desired.  The  idea that right
and left  be on equal footing is expressed as 

$$
 L(\lambda )\,P=P\,L(\lambda ^{-1} ),
\eqno(107)
 $$
 which implies that

$$
 S_-(\lambda ^{-1}  )=S_+(\lambda ),
\eqno(108)
 $$
hence $j'=-j$.  We call $j$ the {\sl Lorentz weight\/} of the solution.  (In
more than one space dimension, it is related to {\sl spin.)\/} 
\skp

The group $\cp^\uparrow$ obtained from $\cp_0$ by adjoining the space
reflection  is called the {\sl orthochronous\/} 
Poincar\'e group in the physics literature, since it still leaves the
direction of time invariant.   The full Poincar\'e group $\cp$  is obtained by
further adjoining {\sl time reversal.\/}   However, the latter must be
antilinear for reasons which need not concern us here  (cf. Streater
and Wightman [28]).  The fact that  $\cp^\uparrow$ leaves the direction of
time invariant implies that the subspaces of positive--and
negative--frequency solutions are invariant under it.  To mix them,
we introduce the following.

\skp

\noindent {\sl Total reflection:\/}  The map $(x,t)\to (-x, -t)$ is another
discrete symmetry of the wave equation, corresponding to $(u,v) \to (-u,
-v)$.  We take the induced mapping on solutions to be

$$
 R:\ f_+(u)\to  f_+(-u),\qquad f_-(v)\to  f_-(-v).
\eqno(109)
 $$

\skp

 Note that unlike translations, Lorentz transformations do not  act
independently on the left-- and right--moving waves.  This will
be remedied by including the next symmetry.
\skp

\noindent{\sl  Dilations:\/} 
Since the wave equation is homogeneous in $x$ and $t$, it is
invariant under  the map $(x,t)\to(\alpha x, \alpha t)$, for
any $\alpha \ne 0$.  Equivalently, $(u,v)\to (\alpha u, \alpha v)$.
It suffices to confine our attention to $\alpha >0$, since dilations with
$\alpha <0$ can be obtained by combining $D(\alpha )$ with $R$.
 As with Lorentz transformations, we shall 
allow a multiplier $M:\,\rr^+\to\cc_*$ in the induced mapping.  Thus $f(x,
t)\to M(\alpha )\,f(\alpha ^{-1} x, \alpha ^{-1} t)$, or

$$
D(\alpha ):\ f_+(u)\to M(\alpha )\,f_+(\alpha ^{-1} u),
\qquad f_-(v)\to M(\alpha )\,f_-(\alpha ^{-1} v).
\eqno(110)
 $$
 In order that dilations form a group, we must have 
$M(\alpha  ^{-1})=M(\alpha )^{-1} $ and $M(\alpha \alpha ')=M(\alpha
)\,M(\alpha ' )$.  Again, continuity implies that $M(\alpha )=\alpha
^{-\kappa} $ for some $\kappa \in\cc$, called  the {\sl conformal
weight\/}  of the solution. 

The following simple argument should convince the reader of the need to
include non--trivial multipliers in eq. (110), \ie to consider conformal
weights other than zero:  Suppose that $f(x,t)$ is a solution with conformal
weight $\kappa $ and let $g(x,t)=(a\partial_x+b\partial_t)\,f(x,t)$, where
$a$ and $b$ are constants.  Then $g$ is also a solution of the wave equation,
and it is easily seen to have conformal weight $\kappa +1$.  In this way we
can {\sl shift\/}  the conformal weight of a solution by any positive integer
$n$ by applying a homogeneous partial differential operator of order $n$
with constant coefficients.  (In three space dimensions, the electromagnetic
potentials satisfy the wave equation in free space;  the electromagnetic
fields are obtained from them by applying first--order partial differential
operators, hence are  solutions but with higher weight.)

Note that dilations commute with
Lorentz transformations.  Composing $L(\lambda )$ and $D(\alpha )$, we
obtain

$$
 f_+(u)\to \alpha ^{-\kappa} \,\lambda ^{-j}\,f_+(\lambda^{-1} \alpha  ^{-1} u),
\qquad f_-(v)\to\alpha^{-\kappa} \,\lambda^j\,f_-(\lambda\alpha^{-1}v).   
\eqno(111)
 $$
In order for the combination of dilations and Lorentz transformation to
act independently on $f_+$ and $f_-$, we must therefore require that
$j=\kappa $.  We shall refer to $\kappa \equiv j\equiv -j'$ simply as the
{\sl weight\/}  of $f$.  Setting $\beta
=\alpha\lambda $ and $\gamma =\alpha/ \lambda $, we then obtain

$$
 f_+(u)\to \beta ^{-\kappa }\,f_+(\beta  ^{-1} u), 
\qquad f_-(v)\to \gamma ^{-\kappa }\, f_-(\gamma  ^{-1} v).
\eqno(112)
 $$
 Consequently, the semi--direct product $\rr^+\times \cp_0$ acts  on $f$ by

$$
 f_+(u)\to  \beta ^{-\kappa }\, f_+(\beta  ^{-1} (u-u_0)), 
\qquad f_-(v)\to  \gamma ^{-\kappa }\, f_-(\gamma ^{-1}(v-v_0)).
\eqno(113)
 $$
This means that {\sl  the group ${\cal G}_0\equiv \rr^+\times \cp_0$
generated by the continuous transformations $T, L$ and $D$ can be
represented as a  direct product $\ca\times\ca$ of two copies  of the affine
group $\ca$ acting independently on the right--moving and left--moving
waves.\/}  We denote by ${\cal G}_1$ the group obtained from ${\cal G}_0$ 
by adjoining $R$, and by ${\cal G}_2$ the group obtained by further 
adjoining $P$. ${\cal G}_1$ contains negative as well as positive dilations, but
the signs of the dilations in the two components are equal.  (They can be
made independent by adjoining yet another discrete symmetry, namely $(x,
t)\to(t, x)$, but we resist the temptation.)  Note  that the subspaces of right--
and left--moving waves are invariant under ${\cal G}_1$ 
but not under ${\cal G}_2$.

Since the affine group is closely related to wavelets, the decomposition
${\cal G}_0\approx \ca\times\ca$ suggests that we apply  separate wavelet
analyses to $f_+$ and $f_-$.  Actually, we shall see that more can be done, 
due to the fact that the wave equation has another, quite unexpected,
symmetry.

\sv2

\bf 4.3. Hilbert Structures on Solutions \rm
\sv1

\noindent So far we have dealt exclusively with solutions in the
space--time domain, where symmetries have a direct and intuitive
meaning.  We now wish to introduce inner products on solutions which
 make the above symmetry transformations unitary.  For this
purpose, it is necessary to venture into the Fourier domain and, later,
 also into the domain of  complex space--time,  invoking the
Analytic--Signal transform introduced earlier.

 \skp

Let   $f(x, t)$ be a solution transforming under ${\cal G}_2$ with weight
$\kappa$.  Representing $f_\pm$ formally by Fourier integrals, we obtain

$$
 f(x, t)=\inr dp\, \epu \fh_+(p)+\inr dp\, \epv \fh_-(p).
\eqno(114)
 $$
The symmetry operations defined earlier can now be represented in
Fourier space.  Translations act  by

$$
 T(u_0, v_0):\ \fh_+(p)\to e^{-2\pi i pu_0 }\, \fh_+(p),
\qquad \fh_-(p)\to e^{-2\pi i pv_0 }\, \fh_-(p),
\eqno(115)
 $$
Lorentz transformations act by
$$
L(\lambda):\ \fh_+(p)\to\lambda^{1-\kappa}\,\fh_+(\lambda p), 
\qquad \fh_-(p)\to  \lambda ^{\kappa-1} \,\fh_-(\lambda^{-1}  p),
\eqno(116)
 $$
space reflection acts by

$$
 P:\   \fh_+(p)\to \fh_-(-p),\qquad 
\fh_-(p) \to  \fh_+(-p),
\eqno(117)
 $$
total reflection acts by

$$
 R:\  \fh_\pm (p)\to  \fh_\pm (-p)
\eqno(118)
 $$
and  dilations act by

$$
 D(\alpha ):\   \fh_\pm (p)\to \alpha ^{1-\kappa }\fh_\pm (\alpha p).
 \eqno(119)
 $$
\skp

Let $s\equiv 2\Re(\kappa )-1\ge 0$ and define $\hs$ to be the the Hilbert
space of all `functions' $g(x)$ such that

$$
 \|g\|_s^2\equiv \inr dp\ |p|^{-s}\,|\hat g(p)|^2<\infty.
\eqno(120)
 $$
In the terminology of Battle [3], this is the  ``massless Sobolev space of
degree $-s/2$.'' The norm of a solution $f(x, t)$ (which may be identified
with the corresponding spinor $\psi \in\cd_s$) is then given by

$$
 \n f\n_s^2\equiv \n \psi \n_s^2=\|f_+\|_s^2+\|f_-\|_s^2.
\eqno(121)
 $$
 From the above actions  it easily follows that the 
transformations  $T, L, R, P$ and $D$ act unitarily on $\cd_s$, thus giving a
unitary representation of ${\cal G}_2$. 
 Denote by $\ch_s^+$ and $\ch_s^-$ the subspaces of $\hs$ with support on
$[0,\infty)$ and $(-\infty, 0]$, respectively. 
(Note that since $s\ge 0$, the value of $\fh_\pm$ at $p=0$ is unimportant; 
when $s=0$, the origin  has zero measure, and when $s>0$, $f_\pm\in\hs$
implies $\fh_\pm(0)=0$.)  The continuous symmetries $T, L$
and $D$ all preserve the  sign  of $p$, hence the group ${\cal G}_0$  generated
by them leaves invariant the subspaces of solutions

$$\eqalign{
\fpp(x,t)&=\int_0^\infty dp\, \epu \fh_+(p)\cr
\fpm(x,t)&=\int_{-\infty}^0 dp\, \epu \fh_+(p)\cr
\fmm(x,t)&=\int_0^\infty dp\, \epv \fh_-(p)\cr
\fmp(x,t)&=\int_{-\infty}^0  dp\, \epv \fh_-(p),
 \cr}
 \eqno(122)
  $$
where the superscipts  denote  positive-- and negative--frequency
components, as in Section 3. 
 The complete solution is $f=\fpp+\fpm+\fmm+\fmp$.
 Note that $\fh_-(p)$ with $p>0$ represents a
{\sl negative--frequency\/}  component  since $v=x-t$.  Thus
$f_+^\pm\in\ch_s^\pm$ but $f_-^\pm\in\ch_s^\mp$. 
The decomposition $\hs=\ch_s^+\oplus\ch_s^-$ therefore gives a 
corresponding decomposition

$$
 \ds=\cd_{s+}^+\oplus\cd_{s+}^-\oplus\cd_{s-}^+\oplus\cd_{s-}^-\,  ,
\eqno(123)
 $$
where $\cd_{s+}^\pm=\ch_s^\pm\oplus\{0\}$ and 
$\cd_{s+}^\pm=\{0\}\oplus\ch_s^\mp$.  Let $\cd_s^\pm=
\cd_{s+}^\pm\oplus\cd_{s-}^\pm$.  The restriction of the
representation of ${\cal G}_2$  to ${\cal G}_0$ leaves invariant all four
subspaces $\ds_\sigma ^\tau $ $(\sigma , \tau =\pm)$, hence it
is reducible.  When $R$ is included, only the subspaces $\ds_+$ and $\ds_-$
remain invariant.  When $P$ is further included, no invariant subspaces
remain and we have an irreducible unitary representation  of ${\cal G}_2$ on
$\ds$.  This shows that the discrete symmetries $P$ and $R$ serve to
`weave' the four representations on $\cd_{s\pm }^\pm,\,\cd_{s\pm }^\mp $ 
(which are associated with the affine group rather than the wave equation)
into a single representation characterising the wave equation as a whole. 
This was the purpose for which they were introduced. 

  However, distinct values of $\kappa $ are not
significantly different at this point.  For  if $f$ has weight $\kappa$, let
$\kappa '\in\cc$ with $ s'\equiv 2\Re(\kappa')-1$ and

$$
\hat g_\pm (p)\equiv |p|^{\kappa '-\kappa}  \,\fh_\pm (p).
\eqno(124)
 $$
Then the corresponding solution $g(x, t)$ has weight $\kappa '$ and,
furthermore, the map $f\to g$ is unitary from $\ds$ onto $\cd_{s'}$.
 This shows that the UIR's with any
two values of $\kappa $ are  unitarily equivalent. If no more could be
said, we might as well restrict ourselves to a single value of $\kappa $,
since solutions of arbitrary  weight can be obtained by the above
method.  (Recall that we could shift the weight {\sl up\/}   by a positive
integer by applying a differential operator;  the above generalizes this
process to arbitrary shifts.)
  Actually, it turns out that for certain special values of $\kappa$,
the group of symmetries can be significantly enlarged, and distinct values of
$\kappa $ then give unitarily inequivalent representations of the larger
group.  Unlike $\ca$, however, the larger group no longer  acts simply on the
Fourier space; instead, its natural domain of action is the {\sl complexified
space--time\/} associated with the AST.  For this reason it becomes
necessary to re--express the norm of $\ds$ in terms of the AST $\ftil$ of
$f$, as will be done next.

\sv2

\bf 4.4. Norms in terms of Analytic Signals \rm
\sv1

\noindent We wish to give  the norms defined above  expressions which are
 local in space, that is, involve only values of $f(x,t)$ at an
arbitrary fixed time $t$, say $t=0$.  Since a knowledge of both $f(x,0)$ and
$\partial_t f(x,0)$ is required to determine $f$, we cannot expect such
expressions to characterize the complete solution $f$.  On the other hand,
the positive-- and negative--frequency parts $f^\pm$ of $f$ satisfy the
first--order {\sl pseudo--differential\/} equation

$$
 -i\partial_t f^\pm=\pm \sqrt{-\partial_x^2}\,f^\pm  \, ,
\eqno(125)
 $$
hence are determined by their initial values $f^\pm(x,0)$. 
(Thus, instead of using $f(x,0)$ and $\partial_t f(x,0)$ as initial data, we use
the symmetric pair $f^\pm(x,0)$.)   We therefore consider separately the
norms on $\ch_s^+$ and $\ch_s^-$, given respectively by

$$\eqalign{
\n  f^+\n_s^2&=\inr dp\, |p|^{-s}\,\left[ \theta (p)\,|\fh_+(p)|^2+
 \theta (-p)\,|\fh_-(p)|^2  \right]   \cr
     &=\int_0^\infty dp\  p^{-s}\left[ |\fh_+(p)|^2+|\fh_-(-p)|^2\right] ,  \cr
\n f^-\n _s^2 &=\inr dp\, |p|^{-s}\,\left[ \theta (-p)\,|\fh_+(p)|^2+
 \theta (p)\,|\fh_-(p)|^2  \right]   \cr
&=\int_0^\infty dp\ p^{-s}\left[ |\fh_+(-p)|^2+|\fh_-(p)|^2\right] .
 \cr}
 \eqno(126)
  $$
The separation of positive and negative frequencies is characteristic of the
AST;  hence let us consider the AST of the complete solution $f$:

$$\eqalign{
 \ftil(x+ix', t+it')&\equiv \inds f(x+\tau x', t+\tau t')\cr
&=\inds\left[f_+(u+\tau u')+f_-(v+\tau v') \right]\cr
&=\ftil_+(u+iu')+\ftil_-(v+iv'),
\cr}
\eqno(127) 
$$
where the imaginary parts of the space--time and light--cone coordinates
are related by $u'=x'+t'$, $v'=x'-t'$ and $\ftil_\pm$ are the AST's of
$f_\pm$.  Recall from Section 3  that the AST's  $\ftil^\pm$ of
$f^\pm(x, t)$ are obtained by restricting $\ftil$  to the forward and backward
tubes $\tb^\pm$, where  $\pm t'>|x'|$.  This suggests that the AST is
well--suited for the analysis of the above norms, since it naturally breaks
$f$ up into its four components.   

The symmetry operations on real space--time extend to the complexified
space--time by $\cc$--linearity, and these extensions induce transformations
on $\ftil$ in the obvious way.  For example, 

$$
 T(u_0, v_0):\ \ftil_+(u+iu')\to\ftil_+(u-u_0+iu'),\quad 
\ftil_-(v+iv')\to\ftil_+(v-v_0+iv').
\eqno(128)
 $$
(Note that the parameters of the symmetries are still real; \eg we do not
consider complex translations.)

Since only the combinations $(x+ix')\pm(t+it')$ enter into $\ftil$, it suffices to
set $x'=0$ and $t=0$.  This is called the {\sl Euclidean region\/}  in quantum
field theory, since the Lorentzian metric $x^2-t^2$ becomes $ x^2+{t'}^2$,
which is Euclidean.  (See Glimm and Jaffe [10].) Note that

$$
 \ftil(x, it')=\ftil_+(x+it')+\ftil_-(x-it')\equiv \ftil_+(z)+\ftil_-(z^*)
\eqno(129)
 $$
is now presented as a sum of  analytic and anti--analytic functions of
$z\equiv x+it'$, as befits a solution of the ``analytically continued'' wave
equation

$$
 \left[{ \partial^2\over \partial x^2}+{ \partial^2\over \partial {t'}^2}\right]
\ftil(x, it')=0,
\eqno(130)
 $$
which states that $\ftil(x, it')$ is harmonic.    In terms of
the Fourier transforms,

$$\eqalign{
\ftil (x, it')=\inr dp\, \left[\theta (pt')\,e^{2\pi ipz} \,\fh_+(p)
+\theta (-pt')\,e^{2\pi ipz^*} \,\fh_-(p)\right] .
 \cr}
 \eqno(131)
  $$
Hence the initial values of $f^\pm$ are given by

$$\eqalign{
F^\pm(x) & \equiv  f^\pm(x, 0)=\lim_{\pm t'\to 0^+}\ftil(x, it')
\equiv \ftil(x, \pm i0) \cr
&=\inr dp\, e^{2\pi ipx}\,\left[ \theta (\pm p)\,\fh_+(p)
+\theta (\mp p)\,\fh_-(p)\right] ,
 \cr}
 \eqno(132)
  $$
from which

$$
 \hat F^\pm(p)=\theta (\pm p)\,\fh_+(p) +\theta (\mp p)\,\fh_-(p).
\eqno(133)
 $$
Returning to the norms, consider first the simplest case $s=0$ (\ie $\kappa
=\h+i\mu$, with $\mu $ real):

$$\eqalign{
\n f^\pm\n _0^2&=\inr dp\, \left[
\theta (\pm p)\,|\fh_+(p)|^2 +\theta (\mp p)\,|\fh_-(p)|^2 \right]  \cr
&=\inr dp\, |\hat F^\pm(p)|^2\cr
&=\inr dx\, |F^\pm(x)|^2
 \cr}
 \eqno(134)
  $$
by Plancherel's theorem.  We have therefore proved

\proclaim Theorem 4. The norms in the positive-- and negative--frequency
subspaces $\cd_0^\pm$ of $\cd_0$ can be be written as
$$\eqalign{
\n f^\pm\n _0^2&=\inr dx\, |\ftil(x, \pm i0)|^2,
 \cr}  
 \eqno(135)
  $$
and the complete norm in $\cd_0$ is therefore
$$
\n f\n _0^2=\inr dx\,\left[  |\ftil(x, i0)|^2+ |\ftil(x, -i0)|^2\right] .\qed
\eqno(136)
 $$

This is not quite local in space since it involves the
boundary values \break\hfill
$\ftil(x, \pm i0)$.   Rather, it is `local' in the 
{\sl  complex\/}  space generated by the AST.  We refer to this property as
{\sl pseudo--locality;\/}   it stands in the same relation to locality as the
pseudo--differential equation (125) stands to the wave equation.  Since $f(x,
0)=\ftil(x, i0)+\ftil(x, -i0)$ and  $\ftil(x, it')$ may be regarded
as a  regularized version of $f(x, 0)$, the term `pseudo--locality' seems
appropriate. 

  \skp

Next, fix $\kappa \in\cc$ with $s>0$.
Can we obtain pseudo--local expressions for the norms with $s>0$? We have

$$\eqalign{
\n f^\pm\n_s^2  &=\inr dp\,|p|^{-s}\,\left[\theta (\pm p)\,|\fh_+(p)|^2+
 \theta (\mp p)\,|\fh_-(p)|^2\right] \cr
&=\int_0^\infty dp\ p^{-s}\,\left[ |\fh_+(\pm p)|^2+ 
|\fh_-(\mp p)|^2\right]  .
 \cr}
 \eqno(137)
  $$
  \proclaim Theorem 5.  For arbitrary  $s>0$, the norms $\n f^\pm\n_s$ in
$\cd_s^\pm$ have the following pseudo--local expressions in terms of the
restrictions of $\ftil$ to the Euclidean region:
$$\eqalign{
\n f^\pm\n _s^2=N_s\inr dx\int _0^\infty dt'\ {t'}^{s-1}\,|\ftil(x, \pm it')|^2,
 \cr}
 \eqno(138)
  $$
where 
$$
 N_s={(4\pi )^s \over \Gamma(s)}.
\eqno(139)
 $$
Hence the norm in $\ds$ is given by
$$
 \n f\n_s^2=N_s\inr dx\inr dt'\, |t'|^{s-1}\,|\ftil(x, it')|^2.
\eqno(140)
 $$

\pf  We prove the theorem for $\n f^+\n _s$. The proof for
$\n f^-\n _s$ is similar.   By eq. (131), $\ftil(x, it')$ can be written as an
inverse Fourier transform 

$$
 \ftil(x, it')=\left[\theta (pt')\, e^{-2\pi pt'}\fh_+(p) +
\theta (-pt')\, e^{2\pi pt'}\fh_-(p)\right]\check{ }\,(x) ,
\eqno(141)
 $$
hence by Plancherel's theorem,

$$\eqalign{
 \inr dx\,|\ftil(x, it')|^2&=\inr dp\, 
\left[\theta (pt')\, e^{-4\pi pt'}|\fh_+(p)|^2 +
\theta (-pt')\, e^{4\pi pt'}|\fh_-(p)|^2\right]  \cr
&=\inr dp\ \theta (pt')\, e^{-4\pi pt'}\left[|\fh_+(p)|^2+|\fh_-(-p)|^2 \right]  .
 \cr}
 \eqno(142)
  $$
When $t'>0$, then $p>0$ in the last integral.
Integrating over $t'$ with the factor ${t'}^{s-1}$, exchanging the order of
integration on the right--hand side and using 

$$
 \int_0^\infty dt'\  {t'}^{s-1}\,e^{-4\pi pt'}=(4\pi p)^{-s}\,\Gamma(s)
=N_s ^{-1}\,p^{-s}
\eqno(143)
 $$
for $s>0$, we obtain

$$\eqalign{
N_s \inr dx \int_0^\infty dt'\,{t'}^{s-1}\,|\ftil(x, it')|^2
&=\int_0^\infty dp\  p^{-s}\,\left[|\fh_+(p)|^2+ |\fh_-(-p)|^2\right]  \cr
&=\n f^+\n _s^2
 \cr}
 \eqno(144)
  $$
as claimed. \qed

\sv2

\bf 4.5.  The Wavelets $e_z$ and their Mother \rm
\sv1

\noindent Here we show that Theorem 5 establishes a resolution of the
identity $I_s$ in $\ds$ ($s>0$) in terms of wavelets $\ez\in\ds$ parametrized
by $z=x+it'$, $t'\ne 0$, in the manner described in Sections 2.2 and 3.3.  
We shall define $\ez$ as an element of $\ds$ whose inner product with
$f\in\ds$ is the {\sl value\/} of the AST $\ftil$  at $(x, it')$, \ie

$$
 \ftil(x, it')=\L\ez, f\R_s\equiv \l\ez_+, f_+\r_s+\l\ez_-, f_-\r_s\,  ,
\eqno(145)
 $$
where $\l\cdot , \cdot \r_s$ and  $\L\cdot, \cdot \R_s$ denote the inner
products in $\hs$ and  $\ds$, respectively.   Since $\ez$ is to belong to
$\ds$, it must itself be a solution.  Eq. (131)  shows that its components in
Fourier space are

$$\eqalign{
\he_+(p)&=\theta (pt')\,|p|^s\,e^{-2\pi ipz^*} \cr
\he_-(p)&=\theta (-pt')\,|p|^s\,e^{-2\pi ipz} .\cr
 \cr}
 \eqno(146)
  $$
(Recall that according to our convention, $\l g, f\r_s$ is anti--linear in $g$.) 
Note that  $\ez_-=e_{z^*+}$ as elements of $\hs$, and that
 $\ez_-$ and $\ez_+$ are orthogonal not only as elements of the
direct sum $\ds=\ds_+\oplus\ds_-$ (since $\ez_\pm\in\ds_\pm$) but also as
elements of $\hs$.  General solutions clearly do not share this property.  The
positive-- and negative--frequency components $e_z^\pm$ of $\ez$ are
obtained by choosing $t'>0$ and $t'<0$, respectively:

$$
 \ez=\cases{e_z^+& if $t'>0$\cr  e_z^-  & if $ t'<0$.\cr}
\eqno(147)
 $$
Hence
 
$$\eqalign{
\hat e_{z+}^+(p)&=\theta (p)\,p^s\,e^{-2\pi ipz^*},\quad t'>0\cr
\hat e_{z-}^+(p)&=\theta (-p)\,|p|^s\,e^{-2\pi ipz},\quad t'>0\cr
\hat e_{z+}^-(p)&=\theta (-p)\,|p|^s\,e^{-2\pi ipz^*},\quad t'<0\cr
\hat e_{z-}^-(p)&=\theta (p)\,p^s\,e^{-2\pi ipz},\quad t'<0.
 \cr}
 \eqno(148)
  $$
For $t'>0$, we have 

$$\eqalign{
\n e_z^+\n _s^2&=\|e_{z+}^+\|_s^2+\|e_{z-}^+\|_s^2  \cr
&=\inr dp\,|p|^s\,\left[ \theta (p)e^{-4\pi pt'}+
 \theta (-p)e^{4\pi pt'}\right]  \cr
&=2\int_0^\infty dp\ p^s\,e^{-4\pi pt'}
={2\Gamma(s+1) \over (4\pi t')^{s+1} }.
 \cr}
 \eqno(149)
  $$
Since $e_{z-}=e_{z^*\!+}$ in $\hs$, we have for $t'<0$

$$
 \n e_z^-\n _s^2={2\Gamma(s+1) \over (-4\pi t')^{s+1} }\,  ,
\eqno(150)
 $$
hence

$$
 \n e_z\n _s^2=\n e_z^+\n _s^2+\n e_z^-\n _s^2=
{2\Gamma(s+1) \over (4\pi |t'|)^{s+1} }\quad \forall t'\ne 0,
\eqno(151)
 $$
since one of the vectors $e_z^\pm$ vanishes.  This proves that $e_z\in\ds$ as
claimed.
\skp

Theorem 5 can be given a  suggestive form in terms of the orthogonal
projection $P(z)$ onto the one--dimensional subspace of $e_z$ in $\ds$. 
For $t'\ne 0$, 

$$\eqalign{
|\ftil(x, it')|^2=\left| \L e_z, f\R_s\right|^2=\n e_z\n _s^2\,\L f, P(z)f\R_s
={2\Gamma(s+1) \over (4\pi |t'|)^{s+1} }\,\L f, P(z)f\R_s\, .
 \cr}
 \eqno(152)
  $$
Substituting this into eq. (140) and simplifying, we obtain

$$
{s \over 2\pi }\inr dx \inr { dt'\over {t'}^2}\,\L f, P(z)f\R_s=\L f, f\R_s\,  .
\eqno(153)
 $$
Upon polarization, this proves the {\sl weak\/}  operator identity

$$
{s \over 2\pi }\inr dx \inr { dt'\over {t'}^2}\,P(z)=I_s \, ,
\eqno(154)
 $$
giving a resolution of the identity in $\ds$ in terms of the projections
$P(z)$.  Note that $dx\,dt'/{t'}^2$ is just the left--invariant measure on the
affine group.  We shall see below that $x$ and $t'$ can indeed be used to
parametrize the affine group, with $t'<0$ corresponding to {\sl negative \/} 
dilations.  The fact that eq. (154)  involves the left--invariant measure on
the affine group shows that the weight function $|t'|^{s-1}$ in the original
measure only served as a normalization factor for $\ez$, so as to make
$\ez_+$ and $\ez_-$ analytic in $z^*$ and $z$, respectively, which in turn
makes $\L\ez_+ ,f\R_s\equiv \ftil_+(z)$ and  $\L\ez_- ,f\R_s\equiv
\ftil_-(z^*)$ analytic in $z$ and $z^*$, respectively.
\skp

The wavelet $e_z$ can be regarded as a {\sl regularization\/}  of the point
$x$ in space, with $|t'|$ as a measure of its diffusion.  Hence $e_z$
transforms naturally under those symmetry operations which leave the
Euclidean region invariant.  These consist of space translations $T(x_0)$,
dilations $D(\alpha )$ and the reflections $P$ and $R$.  Clearly the Euclidean
region is not invariant under Lorentz transformations and time translations. 
(It is not difficult to extend the family of wavelets to one parametrized by
general points $(x+ix', t+it')$ in complex space--time, with $x'\ne \pm t'$. 
This extended family is then invariant under ${\cal G}_2$,  and the above
inner products are obtained by simply integrating over the submanifold
defined by $x'=t=0$.  Then the  invariance of the inner product under the
full group ${\cal G}_2$, known to hold in the Fourier domain, can be made
manifest in the complex space-time domain by introducing conserved
currents and using Stokes' theorem. Cf.  [16], Section 4.5 for a related
treatment of the Klein--Gordon equation.)

 To see how $e_z$ transforms under a symmetry operation, we need only
apply eq. (145).  Thus for  translations,

$$\eqalign{
 \L T(x_0)\,e_z, f\R_s&=\L e_z,T(-x_0) f\R_s=\left(T(-x_0)\,\ftil\right)(x,
it')\cr &=\ftil(x+x_0, it')=\L e_{z+x_0}, f\R_s\,  ,
 \cr}
 \eqno(155)
  $$
hence

$$
 T(x_0)\,e_z=e_{z+x_0}.
\eqno(156)
 $$
Similarly,

$$\eqalign{
\L D(\alpha)\,e_z, f\R_s&=\L e_z, D(\alpha^{-1} )\,f\R_s \cr
&=\alpha ^\kappa \ftil(\alpha x, i\alpha t')
=\alpha^\kappa \L e_{\alpha z}, f\R_s\, ,
 \cr}
 \eqno(157)
  $$
hence

$$
 D(\alpha )\,e_z=\alpha ^{\kappa ^*}e_{\alpha z},\qquad \alpha >0.
\eqno(158)
 $$
Since $P^*=P^{-1} =P$ and $(P\ftil)(x, it')=\ftil(-x, it')$, we have

$$
 Pe_z=e_{-z^*}.
\eqno(159)
 $$
Finally, $R^*=R ^{-1} =R$ and $(R\ftil)(x, it')=\ftil(-x, -it')$ implies

$$
 Re_z=e_{-z}. 
\eqno(160)
 $$
Thus $T(x_0)$ and $D(\alpha )$ generate a single copy of $\ca$ (as opposed
to ${\cal G}_0\approx\ca\times\ca$), and $P$ and $R$ extend this to include space
reflections and negative dilations.  We may therefore begin with a single
{\sl basic\/}  wavelet, say $\phi =e_{i+}\in\ds_+$ (left--moving
wavelet with $z=i$).  Its components in Fourier space are then

$$
 \ph_+(p)=\theta (p)\,p^s\,e^{-2\pi p},\qquad \ph_-(p)\equiv 0.
\eqno(161)
 $$
For $z=x+it'$ with $t'>0$ we obtain all the left--moving wavelets by
applying $T, D$ and $R$ to $\phi $:

$$\eqalign{
 e_{z+}^+&={(t')}^{-\kappa ^*}\,T(x)\,D(t')\,\phi \cr
 e_{z^*+}^-&={(t')}^{-\kappa ^*}\,T(x)\,R\,D(t')\,\phi ,
 \cr}
 \eqno(162)
  $$
and the right--moving wavelets are obtained in the same way from $P\phi
=e_{i-}$.   Note that the choice of $\phi $ depends only on $s$ and not on the
imaginary part of $\kappa $.  We may regard eqs. (162) as providing a {\sl
construction\/} of the UIR with weight $\kappa $ by giving all of its
wavelets.

Modified versions of the left component $\phi_+ $ of $\phi $ have
already appeared in the literature, in connection with the representation 
theory of the affine group.  Namely, if we absorb the Sobolev weight function
$|p|^{-s}$ into the elements $g$ of $\hs$ by defining $\hat
G(p)=|p|^{-s/2}\,\hat g(p)$ (so that $\hat G\in L^2(\rr, dp)$), then $\ph$
becomes

$$
 \hat \Phi_+ (p)=\theta (p)\,p^{s/2}\,e^{-2\pi p},\qquad
\hat\Phi_-(p)\equiv 0.
 \eqno(163)
 $$
 The function $\hat \Phi_+ (p)$ with  $s=1$ first appeared in the classical
papers of Aslaksen and Klauder [1], where it was used as a `fiducial vector'
to obtain a `continuous representation' of the affine group.  Since those
papers contain the first instance of what is now called continuous wavelet
analysis, we suggest that $\phi $ with $s=1$ deserves to be called the {\sl
Mother of all Wavelets.\/}   For $s=1, 2, 3, \cdots,$   $\hat \Phi_+ (p)$ also
appears in the work of Paul [24] in connection with representations of the
affine group and their extension to $\slt$ (see below).
It must be noted, however, that in our case the wavelets are {\sl
prescribed\/}  by the problem at hand (solving the wave equation and
extending the solutions to complex space--time, via the AST)
rather than chosen arbitrarily as a convenient family of functions to be used
in expansions.  This is further discussed in Section 5.

\skp

Having established a resolution of unity in terms of the wavelets $\ez$, let
us now investigate the associated {\sl reconstruction\/}  (cf.  Section
2.2).  The formula corresponding to eq. (22) is

$$
 f(x, t )=N_s\int_{\rr^2} dx_1\, dt_1'\,{|t_1'|}^{s-1}\,e_{x_1+it_1'}(x, t)\,
\ftil(x_1, it_1').
 \eqno(164)
 $$
This gives an expansion of an arbitrary solution $f\in\ds$  in terms of 
the wavelets $e_{z_1}$, this time expressed  in the {\sl real\/} 
space--time domain.  Let us therefore compute the left--moving wavelet
$e_{z_1 +}(x, t)$.   For simplicity,  choose $x_1=0$ and $t_1'>0$; 
the other cases ($x_1\ne 0$, $t_1'<0$ and right--moving wavelets) can be
obtained easily by using $T(x_1), R$ and $P$.  By eq. (146), 

$$\eqalign{
e_{it_1'+}(x, t)=\int_0^\infty dp\ p^s\,e^{2\pi ip(u+it_1')}
={\Gamma(s+1) \over \left(2\pi (t_1'-iu)\right)^{s+1}}.
 \cr}
 \eqno(165)
  $$
Hence

$$
\left| e_{it_1'+}(x, t) \right|^2=
{\Gamma(s+1)^2\over (2\pi)^{2s+2}\, ({t_1'}^2+u^2)^{s+1}}.
\eqno(166)
 $$
At time $t$, this solution is localized in space around $x=-t$, and
its width is proportional to $t_1'$.  As $t_1'\to 0$, the wavelet becomes an
infinitely sharp spike at $x=-t$.  This is not surprising, since $t_1'$ acts as
a scale parameter for the affine group.  Similarly, $e_{x_1+it_1'}(x, t)$ is 
centered near $x=x_1-t$.

\skp

A direct and important interpretation of $t'$ is that
$1/t'$ is proportional to the  average frequency $\nu _{s\pm}$, or {\sl
color,\/}  in the frequency
spectrum of $e_{z\pm}$.  This can be  seen most easily by noting that the
function $p^s\,\exp(-2\pi pt')$ has a maximum at $p=s/(2\pi t')$.  A
more precise argument is based on the quantum mechanical   notion of {\sl
expectation values,\/}  where $|\hat e_{z \pm}(p)|^2$ is viewed as an
(unnormalized) probability distribution for $p$ with respect to the measure
$|p|^{-s}\,dp$. Remembering that the frequency in $\ds_\pm$ is
$\pm p$, this gives

$$\eqalign{
\nu _{s\pm}\equiv 
{\inr |p|^{-s}\,dp\ (\pm p)\, |\hat\ez_\pm(\p)|^2 \over
 \inr |p|^{-s}\,dp\,  |\hat\ez_\pm(\p)|^2} =
\pm\left(-{1 \over 4\pi }{\partial \over \partial t'}\right)
\log\|\ez_\pm\|_s^2= { s+1\over 4\pi t'}.
 \cr}
 \eqno(167)
  $$
(Note that $\nu _{s+}=\nu _{s-}\equiv \nu_s$ and, as expected, $\nu_s$ has
the same sign as $t'$.)   Similarly, one computes the standard deviation in the
frequency to be

$$
 \Delta \nu_s=\sqrt{\left(-{1 \over 4\pi }{\partial \over \partial t'}\right)^2
\log\|\ez_\pm\|_s^2}=
{ \sqrt{s+1}\over 4\pi |t'|}.
\eqno(168)
 $$
Note that eq. (154) assumes an especially simple
form in terms of the  variables $(x, \nu _s)$:

$$
   {{ 2s\over s+1}}\, \inr dx\,\inr d\nu _s\,P\left(x+i{ s+1\over 4\pi \nu
_s}\right) =I_s  \, ,
\eqno(169)
 $$
 which gives {\sl  a resolution of the identity in $\ds$ in terms of the
orthogonal projections onto the wavelet subspaces parametrized by initial
location and color, with all locations and colors given equal  a priori 
probability,\/} since the measure is Lebesgue!
(The appearance of Lebesgue measure is not an accident.  The variables  
$(x,  \nu _s)$ are  {\sl phase--space coordinates,\/} also famous in symplectic
geometry as {\sl Darboux canonical coordinates,\/} and $dx\,d\nu _s$ is 
the corresponding {\sl Liouville measure.\/}  For a similar analysis of
the wavelets associated with the Klein--Gordon equation, see ref.  [16],
Section 4.4.)
\skp

We are now in a position to answer a question which was posed in a more
general context in Section 3.2:   Given a function $F(x,t')$, how can we tell
whether $F=\ftil(x, it')$ for some $f\in\ds$?  Suppose this were the
case for some positive integer $s$. From the polarized version of eq. (140),
\ie

$$
 \L g, f\R_s=N_s\inr dx\inr dt'\, |t'|^{s-1}\, \gtil(x, it')^*\,\ftil(x, it')\,  ,
\eqno(170)
 $$
we obtain by choosing $g=e_{z_1}$ with $z_1=x_1+it_1', z=x+it'\, $:

$$
\ftil(x_1, it_1')=N_s\inr dx\inr dt'\, |t'|^{s-1}\, \L e_{z_1}\, ,\ez\R_s\,\ftil(x,
it')\,  .
\eqno(171)
 $$
The function

$$\eqalign{
K(z_1, z)\equiv \L e_{z_1}\, ,\ez\R_s
 \cr}
 \eqno(172)
  $$
is a {\sl reproducing kernel\/} for the function space 
${\cal F}_s\equiv \{\ftil\st f\in\ds\}$.
 A  computation similar to that for $\n \ez\n_s$  gives

$$
 K(z_1, z)={2\theta (t_1't')\,\Gamma(s+1)\, \over
\left( 2\pi | t_1'+t' |\right)^{s+1} }\,\Re\left[
\left(1-i\,{x_1-x\over t_1'+t' }\right)^{-s-1}\right] . 
\eqno(173)
 $$
The integral operator defined by $K$ acts as the orthogonal projection from
$L^2(|t'|^{s-1}\,dx\,dt')$ to ${\cal F}_s$.  The given function $F$ is therefore
the AST of some $f\in\ds$ if and only if is satisfies the {\sl consistency
condition\/} 

$$
 F(x_1,  t_1')=N_s\inr dx\inr dt'\, |t'|^{s-1}\,K(z_1, z)\,F(x, t')
\eqno(174)
 $$
(cf.  [16], Chapter 1 for an exposition of this general idea).

\sv2

\bf 4.6. Extension to $SL(2,\rr)$ \rm
\sv1

\noindent We are at last ready to extend the symmetry group to a larger
one that will `break' the unitary equivalence of  the UIR's of $\ca$
with different weights.  We do so by showing that for $2\kappa =1, 2, 3,
\cdots$, the unitary action of $\ca$ on each of the subspaces $\ds_\pm$
extends to the group  of real $2\times 2$ matrices of unit determinant, and
that this extension preserves the positive-- and negative--frequency
subspaces $\cd_{s\pm}^+$ and  $\cd_{s\pm}^-$ and is therefore irreducible
when restricted to these subspaces. The extension can be
implemented most easily on the AST's of solutions in $\ds$.  We concentrate
on $\ds_+\approx\hs$ for simplicity and denote its elements by $g$ rather
than $f_+$ to avoid a proliferation of indices.

The action of the affine group on $\hs$ is
induced by the map $u\to \alpha u+\beta $ on $\rr$, which is 
but a very special case of an {\sl order--preserving diffeomophism\/}
$h:\,\rr\to\rr$; \ie  an invertible map $h$ such that both $h$ and $h ^{-1} $ 
are $C^\infty$ and $dh/du>0$.  Under composition, the set of all such maps
forms a group Diff$_+(\rr)$, sometimes called the {\sl pseudo--conformal
group,\/}  of which  $\ca$ is a subgroup.   Note that whereas it takes only two
parameters to specify an affine map, an infinite number of parameters are
needed to specify a general diffeomorphism.  Nevertheless, for $\kappa
=\h$, the   action  of $\ca$ on $\ch_0=L^2(\rr)$ can be extended to
Diff$_+(\rr)$ by defining

$$
 A(h):\  g (u)\to\sqrt{{dh ^{-1}  \over du}}\,g \left(h ^{-1} (u)  \right),
\eqno(175)
 $$
which is unitary  and gives a representation  since
$A(h_1)\,A(h_2)=A(h_1\circ h_2)$.  However, this representation  is
problematic since general elements of Diff$_+(\rr)$ do not preserve the
positive-- and negative--frequency subspaces of $\ch_0$.  To see this, recall
that $g$ is  the boundary--value of its AST $\gtil(z)$ and $\gtil=0$ in
$\cc^\mp$  if $g\in\ch_s^\pm$.  Hence elements of $\ch_s^\pm$ are
boundary--values of functions analytic in $\cc^\pm$. When composed with a
diffeomorphism $h^{-1} $ which is not the boundary--value of a function
analytic  in $\cc^\pm$, the result cannot be the boundary--value of a function
analytic in $\cc^\pm$.   (Diff$_+(\rr)$ plays important roles in string theory
and quantum field theory;  there, the fact that it mixes positive and negative
frequencies leads to a difficulty known as  the ``conformal anomaly.'') Since
the decomposition into positive-- and negative--frequency components is
fundamental to our approach, the representation theory of  Diff$_+(\rr)$ will
not be pursued further.  However,  the above discussion suggests that we
confine ourselves to  diffeomorphisms of $\rr$ which extend `naturally' to
$\cc$ and  whose restrictions to  $\cc^\pm$ are {\sl holomorphisms\/} 
(analytic diffeomorphisms) of $\cc^\pm$.  That is, we consider symmetry
operations on $\hs$ which are induced from geometric maps of the complex
space $\cc^+\cup\cc^-$ associated with the AST rather than geometric maps
of the real space $\rr$. Since  only  the analytic functions $\gtil(z)$  or their 
boundary--values $\gtil(x\pm i0)$ enter into the pseudo--local inner
products of $\hs$, we might even get away with maps which have
singularities on $\rr$.  Thus we look for the group of  holomorphisms of
$\cc^\pm$. Now it is well--known that the group of holomorphisms of the
Riemann sphere $\rs\equiv \cc\cup\{\infty\}$  is $SL(2, \cc)$, the set of all
complex $2\times 2$ matrices with unit determinant, acting by
fractional--linear (M\"obius) transformations.  The subgroup preserving
$\cc^\pm$, hence also the one--point compactification $\rh\equiv
\rr\cup\{\infty\}$, then consists of all such {\sl real\/}  matrices, and is
denoted $\slt$.  We shall write $G\equiv \slt$ throughout this subsection. 
An element of $G$, being a diffeomorphism of $\rh$ rather than $\rr$, will 
have a singulariy in $\rr$ (the inverse image of $\infty$) unless it has
$\infty$ as a fixed point.

\skp

We begin by constructing two inequivalent UIR's of $G$ which extend the
UIR's of $\ca$ on $\ch_0^\pm$.  The matrix 

$$
 \sigma =\left(\matrix{a&b\cr c&d\cr}\right),\qquad ad-bc=1,
\eqno(176)
 $$
acts on $\cc^+\cup\cc^-=\rs\backslash\rh$ by the fractional--linear
transformation

$$
 \sigma :\ z\to\sigma (z)\equiv {az+b \over cz+d}\equiv w.
\eqno(177)
 $$
(This  action will be derived below.)  From the easily derived identity

$$
 w-w^*={z-z^* \over |cz+d|^2}
\eqno(178)
 $$
it follows that $\sigma $ leaves $\cc^\pm$ invariant.
Note that the boundary map $\sigma (u) $ on $\rr$ has a singularity at
$u=-d/c$, in agreement with the above discussion.  Since
$d\sigma(u)/du=(cu+d)^{-2}$,   the induced map on boundary--values in
$\ch_0$ is, according to eq. (175),

$$
 A(\sigma ^{-1} ):\ g (u)\to |cu+d| ^{-1} g \left({au+b \over cu+d}\right).
\eqno(179)
 $$
The singularity at $u=-d/c$ poses no problem, since  $C^\infty$
functions of compact support $g \in \cd(\rr)$ vanish at infinity and
$\cd(\rr)$ is dense in $\ch_0$.  The set of $\sigma $'s whose boundary maps
are non--singular is, in fact, just the affine group $\ca$, since $c=0$ implies

$$
 \sigma (u)=a^2u+ab\equiv \alpha u+\beta,
\eqno(180)
 $$
giving the relation between the parameters of $\ca$ and those of $G$.  
To represent $A(\sigma ^{-1} )$ on $\gtil(z)$, we first replace the multiplier
$|cu+d| ^{-1}$ with $(cu+d) ^{-1}$, which also leads to a representation and,
unlike the former, can be extended analytically. 
This does not affect the norm as $z\to u\in\rr$. Thus define

$$
  B(\sigma^{-1} ):\ \tilde g (z)\to (cz+d) ^{-1} \tilde g \left({az+b \over
cz+d}\right). 
\eqno(181)
 $$
 For $z\in\cc^\pm$, we obtain the actions on $\ch_0^\pm$, which are 
unitary by the same argument as used for Diff$_+(\rr)$.  This gives
inequivalent UIR's of $\slt$ on $\ch_0^\pm$.
  Comparison with Bargmann's classification
 shows that these  coincide with the two representations of
the ``mock discrete series'' with $s=0$ (Lang [20], pp. 120 and 123).

\skp

We can arrive at the action of $G$ on $\cc^\pm$ (eq. (177)) by the method of
cosets.  Consider the subroups $T$ and $H$ of $G$ given by

$$\eqalign{
 T&=\left\{t\,(u)\equiv \left( \matrix{ 1&u\cr 0&1\cr} \right)
\st u\in\rr\right\}\cr
 H&=\left\{h(c, d)\equiv \left( \matrix{ d ^{-1} &0\cr c& d \cr} \right)
\st d\ne 0\right\}. 
 \cr}
 \eqno(182)
  $$Then $T$ is isomorphic to $\rr$, and $H$ is isomorphic to $\ca$. 
Moreover, any element of $G$ with $d\ne 0$ can be written uniquely in the
form

$$
 \sigma \equiv \left(\matrix{a&b\cr c&d\cr}\right)=
\left(\matrix{1&b/d\cr 0&1 \cr}\right) 
  \left(\matrix{d ^{-1} &0\cr c&d \cr}\right) =t(b/d)\,h(c, d).
\eqno(183)
 $$
  Hence we may coordinatize $G$ (except for a set of Haar
measure zero) by the product of sets $TH$. (This does not mean that $G$ is a
direct product since $T$ and $H$, as subgroups of $G$, do not commute.) 
The space $U\equiv G/H$ of right cosets is then parametrized by $u\in\rh$,
as follows:  If $d\ne 0$, then

$$
 \sigma H\equiv \left\{\sigma k\st k\in H \right\}=
\left\{ t\,(b/d) \,h(c,d)\,k\st k\in H\right\}=t\,(b/d)\,H\equiv H_{b/d}.
\eqno(184)
 $$
The set of  matrices with $d=0$ forms a single coset, which will
be denoted by $H_\infty$.  Thus $U\approx \rh$.  Now $G$
acts on $U$ by left multiplication, and this translates into an action of $G$ on
$\rh$ as follows:  If $u\ne\infty$, then

$$
\left(\matrix{a&b\cr c&d\cr}\right) H_u=
\left(\matrix{a&b\cr c&d\cr}\right)\left(\matrix{1&u\cr 0&1 \cr}\right)H =
\left(\matrix{a&au+b\cr c&cu+d\cr}\right)H=H_{\sigma (u)}, 
\eqno(185)
 $$
with $\sigma(u)$ as given by eq. (177).  The action on $H_\infty$ is
similarly found to be

$$
 \left(\matrix{a&b\cr c&d\cr}\right) H_\infty=H_{a/c}.
\eqno(186)
 $$
(If $c=0$, then $H_\infty$ is invariant under $\sigma $.)

Thus we see that the action of $G$ on $\rr$ (more precisely, on $\rh$) can
be derived from the action of $G$ on itself.  Similarly, the 
representation of $G$ on $\lt$ obtained above can be viewed as a
group--theoretical construction:  The Lebesgue measure $du$ on $\rr$ can
be extended to $U=\rh$ by letting $\{\infty\}$ have measure zero; this
extension $d\hat u$ is quasi--invariant under the action of $G$, and
functions in  $\lt=L^2(U,d\hat u)$ may be interpreted as 
functions on $G$ which are constant on each coset $H_u$. 
The above representation can then be interpreted
as being ``induced'' (Lang [20], chapter 3) from the identity representation  of
$H$.  Since $H$ is not compact, non--zero functions on $G$ which are constant
on the cosets $H_u$ are not square--integrable on $G$.  This proves that the
above representation is not  square--integrable on $G$.  That means that we
cannot use it to obtain a resolution of the identity in $L^2(G)$. 

\skp

We now construct some other  representations of $G$, with weights
$\kappa $ other than $\h$, which {\sl do\/}  turn out to be 
square--integrable on $G$. 
 Let $\kappa $ be real, so that $s=2\kappa -1$.   Define the action of $G$ on
$\hs$ by

$$
B(\sigma ^{-1} ):\ \tilde g(z )\to (cz +d)^{-s-1}\,\tilde g\left(
{ az+b\over cz+d}\right).
\eqno(187)
 $$
We must have $s\in\ZZ$ in order to preserve analyticity.  For
$\kappa =\h$, this coincides with the earlier action.  Note that for
$a=1/\sqrt{\alpha},\,b=-\beta/\sqrt{\alpha},\, c=0$ and $d=\sqrt{\alpha}$
we obtain

$$
 B(\sigma ^{-1} ):\ 
  \tilde g(z)\to \alpha ^{-\kappa }\,\tilde g(\alpha ^{-1}(z-\beta ))\,  ,
\eqno(188)
 $$
which shows that the action restricts to that of the affine group on $\hs$. 

\skp
To show that $B(\sigma ^{-1} )$ acts unitarily, we need to prove that it
preserves the pseudo--local norm introduced in Theorem 5.  There we found
that for $G$ replaced by its subgoup $\ca$, we needed $s>0$.  Since $s$ must 
now, in addition, be an integer, we therefore have $2\kappa -1=s=1, 2, 3,
\cdots$. We must show that 

$$\eqalign{
\int_\cc d^2z\,|\Im(z)|^{s-1}\,  |cz+d|^{-2s-2} \left|  
\gtil \left({ az+b\over cz+d}\right) \right|^2 
=\int_\cc d^2z\,|\Im(z)|^{s-1}\,  \left| \gtil(z) \right|^2 \cr},
 \eqno(189)
  $$
where $d^2z$ denotes Lebesgue measure in $\cc$.
Let $w=\sigma (z)=(az+b)/(cz+d)$.  Then $dw/dz=(cz+d)^{-2}$, hence
$$
 d^2w=|cz+d|^{-4}\,d^2z,\quad \hbox{and}\quad \Im(w)=|cz+d|^{-2}\Im(z),
\eqno(190)
 $$
where the second equality follows from eq. (178).  Thus

$$\eqalign{
\int_\cc d^2z\,|\Im(z)|^{s-1}\,|cz+d|^{-2s-2}\,|\gtil(w)|^2
=\int_\cc d^2w\,|\Im(w)|^{s-1}\,|\gtil(w)|^2=\|g\|_s^2,
 \cr}
 \eqno(191)
  $$
proving  the result.  The unitary representation  on $\hs$ decomposes into a
direct sum of UIR's on $\ch_s^\pm$.  All these representations are
inequivalent, and they are known collectively as the {\sl discrete
series\/} (Gelfand et al. [9], Lang [20]).  
\skp

The action of $\slt$ on the wavelets $\ez_+\in\ds_+\approx\hs$ is easily
computed.  By the unitarity of $B(\sigma )$, we have

$$\eqalign{
\l B(\sigma )\,\ez_+, g\r_s &=\l\ez_+, B(\sigma ^{-1} )\,g\r_s
=\left(B(\sigma ^{-1} )\,\gtil\right)(z)\cr
&=(cz+d)^{-s-1}\,\gtil(\sigma (z))=(cz+d)^{-s-1}\,\l e_{\sigma (z)+}, g\r_s,
 \cr}
 \eqno(192)
  $$
hence 

$$
 B(\sigma )\,\ez_+=(cz^*+d)^{-s-1}\,e_{\sigma (z)+} \,  .
\eqno(193)
 $$

\skp

The representations of the discrete series are square--\-integrable over
$G$. The subgroup of $G$ which leaves $z=\pm i$ invariant is $SO(2)$, hence
$\cc^\pm\approx G/SO(2)$.  Since $SO(2)$ is compact (unlike its
counterpart $H$ for $s=0$), the norms in $\ch_s^\pm$ can be
rewritten as integrals over all of $G$ rather than just the homogeneous
spaces $\cc^\pm$.  

\skp

Returning to the wave equation, we obtain mutually inequivalent  UIR's  of
what we shall call the {\sl restricted conformal group\/} \/ ${\cal
C}_0\equiv G\times G$ on $\cd_{s+}^+ \, , \cd_{s+}^-\, ,\cd_{s-}^-$ and
$\cd_{s-}^+$  $(s=0, 1, 2, \cdots)$. When the total reflection $R$ is included,
we obtain inequivalent UIR's of the resulting group \/ ${\cal C}_1$ on
$\ds_+$ and $\ds_-$.  When  the space reflection $P$ is further included, we
obtain  a single set of mutually inequivalent  UIR's of the resulting group \/
${\cal C}_2$ on $\ds$.
  As they did for ${\cal G}_0$, the reflections unify the
four subspaces  $\cd_{s\pm}^\pm$  into a single one representing the wave
equation as a whole.

\VE

\cl{\bf 5. Concluding Remarks:  Dedicated Wavelets} 
\sv1

\noindent  The wavelet analysis developed  in [16] for the
Klein--Gordon equation has the interesting feature
that the wavelets  are ``dedicated'' to the equation rather than
being merely a convenient set of functions to be used in expansions.  (This is
somewhat reminiscent of the situation in the spectral theorem, where
expansions are customized to a given operator.)  The reward for such
dedication is that symmetry operations (such as translations, rotations,
Lorentz transformations, and even time evolution) take wavelets to wavelets. 
This has the practical consequence of making the description economical and
precise.   For example,  wavelets obtained by taking tensor products of
one--dimensional wavelets cannot be rotated; consequently, a function
consisting of but a few wavelets in one coordinate system is represented
(inefficiently) by a combination of many wavelets in a rotated coordinate
system.  Similar considerations apply to the dedicated wavelets associated
with the wave equation in $\rr^2$  developed in the last section.  But since
there is now only one space dimension, the results are somewhat less
dramatic: There are only two directions, left and 
right, giving rise to the labeling $\ez_+$ and $\ez_-$.  We believe that the
results of Section 4 generalize to $d> 1$ space dimensions, where the set of
directions is parametrized by $S^{d-1}$.  (The case $d=1$ is degenerate since
$S^0=\{\pm 1\}$ is disconnected.) Work on this is in progress.

\sv4

\cl{\bf References}
\sv2

\item{[1]} Aslaksen,~E.~W.  and Klauder,~J.~R., Unitary representations of the
affine group, \JMP {\bf 9 } (1968), 206-211;
Continuous representation  theory
using the affine group, \JMP {\bf 10 } (1969), 2267-2275.
\sv1
\item{[2]}  Battle,~G.,  A block spin construction of ondelettes.  Part I:
Lemari\'e functions,  Communications in Mathematical Physics 
{\bf 110} (1987), 601-615.
\sv1
\item{[3]}  Battle,~G., Wavelets: A renormalization point of view, in {\sl
Wavelets and Their Applications,\/} G.~Beylkin, R.~R.~Coifman, 
I.~Daubechies, S.~Mallat, Y.~Meyer, L.~A.~Raphael and M.~B.~Ruskai
(eds.),    Jones and Bartlett, to appear.
\sv1
\item{[4]} Born,~M. and Wolf,~E., {\sl Principles of Optics,\/} fifth edition,
Pergamon Press, Oxford, 1975.
\sv1
\item{[5]} Daubechies,~I., Grossmann, A. and Meyer, Y., Painless 
nonorthogonal  expansions, \JMP {\bf 27 } (1986), 1271-1283.
\sv1
\item{[6]} Daubechies,~I., Orthonormal bases of compactly supported 
wavelets, Communications on  Pure and Applied Mathematics {\bf 41 }
(1988), 909-996.
\sv1
\item{[7]} Daubechies,~I., {\sl Wavelets,\/} Lecture notes of NSF/CBMS Regional
Conference at the Univ. of Lowell,  SIAM, 1992, to appear.
\sv1
\item{[8]} Gabor,~D., Theory of communications, J. Inst. Elec. Eng. {\bf 93 }
(1946), 429-457.
\sv1
\item{[9]} Gelfand,~I.~M.,  Graev,~M.~I. and Vilenkin,~N.~Ya., {\sl
Generalized Functions,\/}  vol. 5, Academic Press, New York, 1966.
\sv1
\item{[10]} Glimm,~J. and Jaffe,~A., {\sl Quantum Physics: A Functional Integral
Point of View,\/} second edition, Springer, New York, 1987.
\sv1
\item{[11]} Helgason,~S., {\sl Groups and Geometric Analysis,\/}  
Academic Press, New York, 1984.
\sv1
\item{[12]}  Holschneider,~M., Inverse Radon transforms through inverse
wavelet transforms, preprint, CNRS--Luminy, 1990.
\sv1
\item{[13]} Kaiser,~G., Phase--Space Approach to Relativistic Quantum 
Mechanics, Ph. D. Thesis, Mathematics Department, University of 
Toronto, 1977.
\sv1
\item{[14]} Kaiser,~G, Phase--space approach to relativistic quantum 
mechanics.  Part I: Coherent--state representation  of the Poincar\'e
group, \JMP {\bf 18} (1977), 952-959;
part II: Geometrical aspects, \JMP {\bf 19} (1978), 502-507;
part III: Quantization, relativity, localization and gauge freedom,   \JMP
{\bf 22} (1981), 705-714.
\sv1
\item{[15]} Kaiser,~G., Quantized fields in complex spacetime, Annals of 
Physics {\bf 173} (1987), 338-354.
\sv1
\item{[16]} Kaiser,~G., {\sl Quantum Physics, Relativity, and Complex
Spacetime: Towards a New Synthesis,\/} 
 North--Holland,  Amsterdam, 1990.
\sv1
\item{[17]} Kaiser,~G., Generalized wavelet transforms. Part  I: The windowed 
X--Ray transform, Technical Reports Series \#18, University of Lowell,
1990;  \hfill\break
part  II: The multivariate analytic--signal transform, Technical Reports
Series \#19, University of Lowell, 1990.
\sv1
\item{[18]} Kaiser,~G., An algebraic theory of wavelets. Part I: Operational
calculus and complex structure, \sl SIAM J. Math. Anal. \rm {\bf 23},  \# 1 (1992), to appear.
\sv1
\item{[19]} Klauder,~J.~R. and Sudarshan,  E. C. G., {\sl Fundamentals of
Quantum Optics,\/} Benjamin, New York, 1968.
\sv1
\item{[20]} Lang,~S., $SL_2(\rr)$, Springer, New York, 1985.
\sv1
\item{[21]} Lemari\'e,~P., Ondelettes \`a localisation exponentielle,
 J. Math. Pures et Appl. {\bf 67 } (1988), 227-236.
\sv1
\item{[22]} Meyer,~Y. S\'eminaire Bourbaki {\bf 38 } (1985-86), 662.
\sv1
\item{[23]} Meyer,~Y., {\sl Ondelettes et Op\'erateurs,\/}  vols. I, II, Hermann,
Paris, 1990.
\sv1
\item{[24]}Paul,~T., Functions analytic on the half--plane as quantum
mechanical states, \JMP {\bf 25 } (1984), 3252-3263.
\sv1
\item{[25} Rudin,~W., {\sl Fourier Analysis on Groups,\/}  Interscience, New
York, 1960.
\sv1
\item{[26]} Stein,~E., {\sl Singular Integrals and Differentiability Properties
of Functions,\/} Princeton University Press, Princeton, 1970.
\sv1
\item{[27]} Stein,~E. and~Weiss, G., {\sl Introduction to Fourier Analysis on
Euclidean Spaces,\/}  Princeton University Press, Princeton, 1971.
\sv1
\item{[28]} Streater,~R.~F. and Wightman,~A.~S., {\sl PCT, Spin \& Statistics,
And All That,\/} Benjamin, New York, 1964.

\bigskip

\bye